\documentclass[prb,showpacs,twocolumn,floatfix]{revtex4}
\usepackage{graphicx}
\begin{document}
\def\omegav{{\mbox{\boldmath{$\omega$}}}}
\def\sigmav{{\mbox{\boldmath{$\sigma$}}}}
\def\tauv{{\mbox{\boldmath{$\tau$}}}}
\def\rhov{{\mbox{\boldmath{$\rho$}}}}
\def\deltav{{\mbox{\boldmath{$\delta$}}}}
\def\chiv{{\mbox{\boldmath{$\chi$}}}}
\def\muv{{\mbox{\boldmath{$\mu$}}}}
\def\oh{{\scriptsize 1 \over \scriptsize 2}}
\def\of{{\scriptsize 1 \over \scriptsize 4}}
\def\tf{{\scriptsize 3 \over \scriptsize 4}}
\title{Theoretical Analysis of the ``Double-q'' Magnetic Structure
of CeAl$_2$}

\author{A. B. Harris,$^1$ and J. Schweizer$^2$}

\affiliation{(1) Department of Physics and Astronomy, University
of Pennsylvania, Philadelphia, PA, 19104\\(2)CEA-Grenoble, DSM/DRFMC/SPSMS/MDN,
38054 Grenoble Cedex 9, France}
\date{\today}

\begin{abstract}
A model involving competing short-range isotropic Heisenberg interactions
is developed to explain the ``double-q'' magnetic structure of CeAl$_2$.
For suitably chosen interactions terms in the Landau expansion
quadratic in the order parameters explain the condensation of
incommensurate order at wavevectors in the star of
$(1/2-\delta, 1/2+ \delta, 1/2)(2 \pi/a)$, where $a$ is the cubic lattice
constant. We show that the fourth order terms in the Landau expansion
lead to the formation of the so-called ``double-q'' magnetic structure
in which long-range order develops simultaneously at two
symmetry-related wavevectors, in striking agreement with the
magnetic structure determinations. Based on the value of the ordering
temperature and of the Curie-Weiss $\Theta$ of the susceptibility, we
estimate that the nearest neighbor interaction, $K_0$,
is ferromagnetic with $K_0/k=-11\pm 1$K and the next-nearest neighbor
interaction $J$ is antiferromagnetic with $J/k=6\pm 2$K.  We also briefly
comment on the analogous phenomena seen in the similar system, TmS.
\end{abstract}
\pacs{75.25.+z, 75.10.Jm, 75.10.Dg} 
\maketitle

\section{INTRODUCTION}

CeAl$_2$ (CEAL) is a metallic system whose magnetic
structure has been the object of some controversy for several
years.\cite{CEAL1,CEAL2,CEAL3,CEAL4,CEAL5,CEAL6}
Initial studies\cite{CEAL1,CEAL2} indicated the existence of
incommensurate long-range magnetic order on the Ce ions with a single
wavevector in the star of ${\bf q}$, where
\begin{eqnarray}
{\bf q} &=& (2 \pi /a) (1/2-\delta, 1/2+\delta, 1/2) \equiv (2 \pi /a) 
\hat {\bf q} \ ,
\label{qeq} \end{eqnarray}
with $\delta=0.11.$\cite{CEAL1,CEAL2}
Later\cite{CEAL3} it was proposed that this structure involved the
simultaneous condensation of three wavevectors in the star of ${\bf q}$,
but this suggestion of a ``triple q" structure was refuted in
Ref. \onlinecite{CEAL4}.  More recent work\cite{CEAL5,CEAL6}
showed that the structure was in fact a ``double q" one
in which exactly two wavevectors in the star of ${\bf q}$ were
simultaneously condensed. In addition, continued interest in CEAL
is due to its Kondo-like behavior.  Initial indications of this came
from the observation of a minimum in the resistivity at about
15K, which was attributed to spin compensation.\cite{RES0}
A single impurity model, with a Ce$^{3+}$ ion in a cubic crystal field
interacting with the conduction band, was able to account for most of
the electrical properties.\cite{RES2} Moreover, when, in neutron experiments,
no third order magnetic satellite appeared at low temperature, the Kondo
effect was invoked to explain why the moment of a Kramers ion did not
saturate in the zero-temperature limit.\cite{CEAL2}  This objection
is partially removed by the double-q structure.\cite{CEAL6}
Moreover, an analysis\cite{OHKAWA} of multi-q states claims that 
the double-q structure can not be explained if CEAL is regarded as
an itinerant-electron magnet.

In this paper we proceed under the assumption
that although Kondo effects may be present due to the coupling of the
Ce 4f electron to the conduction band, the magnetic structure can
be understood in terms of interaction between localized moments on
the Ce ions.  Since the lattice structure is fcc,
it is apparent that antiferromagnetic interactions between
shells of near neighbors could compete and might then explain the
incommensurability.  However, no concrete calculations of this
type have yet appeared.  It is also interesting that this
system does not follow the simplest scenario for incommensurate
magnets,\cite{Nagamiya} namely, as the temperature is lowered,
a phase transition occurs in which a modulated phase appears
with spins confined to an easy axis, and then, at a lower temperature
a second phase transition occurs in which transverse order
develops, so as to partially satisfy the fixed length spin constraint
expected to progressively dominate as the temperature is lowered.  
Instead, in CEAL, there is no second phase transition, and in the
ordered phase one has the simultaneous condensation of long-range order
at two symmetry-related wavevectors.\cite{CEAL5,CEAL6} There are
two aspects of this behavior that have not yet been explained.
1) The incommensurate wavevector lies close to, but not exactly along
the high symmetry $(1,1,1)$ direction and 2) although
so-called ``triple-q'' systems are well known,\cite{3qa,3qb,3qc,3qd,3qe}
in which the incommensurate ordered state consists of the simultaneous 
superposition of three wavevectors, it is unusual, in a cubic
system, to have a ``double-q'' state\cite{dq1,dq2} consisting of the
simultaneous superposition of exactly two wavevectors.

The aim of this paper is to develop a model which can explain the
above two puzzling features.  We first address the determination of
the incommensurate wavevector. Some time ago, Yamamoto and Nagamiya\cite{YN}
(YN) studied the ground state of a simple fcc antiferromagnet with
isotropic nearest-neighbor (nn) and next-nearest neighbor (nnn)
Heisenberg interactions and found a rich phase diagram in terms of
these interactions whose coupling constants we will denote here as
$J$ and $M$, respectively.  We perform an equivalent calculation for a
related model appropriate to CEAL based on an analysis of the terms in
the Landau expansion of the free energy in the paramagnetic phase.
By studying the instability of this quadratic form which occurs as
the temperature is lowered, one can predict the magnetic structure
of the ordered phase.  In particular, one can thereby determine the
wavevector at which this instability first occurs.  This phenomenon
is referred to as ``wavevector selection." As Nagamiya's
review\cite{Nagamiya} indicates,  correct wavevector selection in
CEAL must require a model which involves competition between
nn and further neighbor interactions.  For the fcc structure of CEAL
the most convenient model which almost explains wavevector selection
involves nn, nnn, and fourth-neighbor interactions.  Based
on our insight developed from this model, we suggest the how more
general interactions can completely explain wavevector selection.
Although we invoke more distant than nn interactions, the magnitudes
of the further neighbor couplings needed to explain the nonsymmetric
wavevector of CEAL decrease with increasing separation and are reasonable, 
especially in view of the possibility of Ruderman-Kittel-Kasuya-Yosida
(RKKY)\cite{RK}
interactions in this metallic system.  Because our main interest
lies in explaining wavevector selection, we have completely ignored
anisotropy, whose major effect is to break rotational
invariance and select spin orientations.  Coincidentally we note several
regions in parameter space for these models in which one has a
multiphase point (at which wavevector selection is incomplete).
This phenomenon is perhaps most celebrated in the Kagom\'{e}\cite{KAG}
and pyrochlore\cite{PYR} systems.  Based on these results we also
point out that there are likewise regions of parameter space
that could explain wavevector selection\cite{TMS1,TMS2,TMS3}
in the similar Kondo-like system TmS.

The second stage of our calculation for CEAL involves an analysis
of the fourth order terms in the Landau expansion, because it is
these terms which dictate whether only one or more than one
wavevector in the star of ${\bf q}$ is simultaneously 
condensed to form the ordered phase.  For this analysis there
are two plausible ways to proceed.  An oft-used approach\cite{dq2} is
to determine the most general fourth order term allowed by symmetry
and then see whether some choice of allowed parameters can explain
a ``double q" state.  The virtue of this method is that it corresponds to
the use of fluctuation-renormalized mean-field theory.
A drawback, however, is
that it is hard to know whether the allowed parameters are appropriate
for the actual system.  Here we adopt a contrary procedure in which
only the ``bare" (unrenormalized) fourth order terms are considered.
Obviously, these terms do have the correct symmetry, and although
they might not be the most general possible terms, they do ensure that
the values of the parameters are plausible.

The organization of this paper follows the above plan.  In Sec. II we extend
the analysis of YN to fcc magnets with three shells of isotropic exchange
interactions, but even this model only partially explains the wavevector
selection seen in CEAL.  In Sec. III we invoke more distant interactions,
whose existence is attributed to either RKKY interactions,\cite{RK}
or indirect interactions via excited crystal field states, as discussed
in Appendix C.  Thereby we explain wavevector selection in CEAL and also
in the similar system TmS.  Here we also use the observed ordering
temperature and data for the zero wavevector susceptibility to estimate
values of the dominant exchange interactions.  In Sec. IV
we analyze the fourth order terms in the Landau expansion and show that
they naturally lead to the ``double-q'' state observed\cite{CEAL5,CEAL6}
in CEAL. Our results are briefly summarized in Sec. V.

\section{WAVEVECTOR SELECTION FOR A "3-$J$" MODEL}

In isotropic Heisenberg models of magnetic systems
with only nearest neighbor (nn) interactions on, say, a
simple cubic lattice, the magnetic structure of the ordered phase
is trivially constructed if the sign of the interaction is known.
In more complicated models it may happen that next-nearest neighbor (nnn)
interactions compete with the nn interactions, in which case
the magnetic structure may be an incommensurate one.\cite{Nagamiya}  In
this case, the quadratic terms in Landau free energy (which we study
below) will be such that, as the temperature is lowered, the paramagnetic
phase develops an instability, relative to the development of
long-range magnetic order, at a wavevector ${\bf q}$ (or more properly,
at the star of ${\bf q}$).  For CEAL our aim is to study this
``wavevector selection," and explain how a model of exchange
interactions can lead to the observed ordering wavevectors.

\begin{figure}
\begin{center}
\includegraphics[width=9cm]{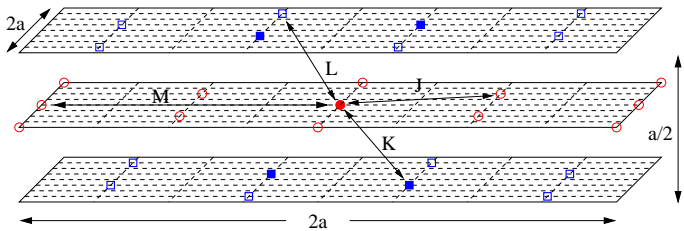} 
\caption{\label{3J} (Color online) Exchange interactions for CEAL, showing
only the magnetic Ce atoms in three (001) planes.  The structure can be thought
of as consisting of two interpenetrating fcc sublattices.  The Ce
$\tauv_1$ atoms are circles (red online) and the Ce $\tauv_2$ atoms
are squares (blue online). Filled squares represent the four
Ce $\tauv_2$ atoms which are nearest neighbors to the Ce $\tauv_1$
atom at the origin, indicated by a filled circle. The exchange constants
between nn's, nnn's, third neighbors, and fourth neighbors are $K$, $J$,
$L$, and $M$, respectively.}  
\end{center}
\end{figure}

For this purpose, this section is devoted to an analysis of the
quadratic terms in the free energy which determine wavevector 
selection. We first note that the space group of CEAL is
Fd$\overline 3$m (space group \#227 in Ref. \onlinecite{HAHN})
which is an fcc system with two
Ce atoms per fcc unit cell at locations
\begin{eqnarray}
\tauv_1 = (0,0,0)\ , \ \ 
\tauv_2 = (1,1,1)(a/4)\ .
\end{eqnarray}
This means that each Ce ion has a tetrahedron of
nn's and we will treat further neighbor interactions as in an
fcc Bravais lattice.  Note that the two sites at $\tauv_1$ and $\tauv_2$
are related by inversion symmetry relative to the point $(1,1,1)(a/8)$.
We introduce the following simple model of exchange interactions,
\begin{eqnarray}
{\cal H} &=& \sum_{{\bf R}, n ; {\bf R}', n'}
J_{n,n'}^{(0)}({\bf R}, {\bf R}')
{\bf S}_{\rm op}({\bf R}+\tauv_n) \cdot {\bf S}_{\rm op}
({\bf R}'+\tauv_{n'} ) \ ,  \label{HAM} \end{eqnarray}
where ${\bf S}_{\rm op}({\bf R}+\tauv_n)$ is the spin operator at
${\bf R}+\tauv_n$.  Here we treat the model having three shell
of interactions, so that the only nonzero $J$'s are
\begin{eqnarray}
J_{12}^{(0)}({\bf R},{\bf R}') &=& J_{21}^{(0)}{\bf R}',{\bf R})
\nonumber \\ &=&K^{(0)}
\ \ {\rm if} \ |{\bf R} + \tauv_1 - {\bf R}'- \tauv_2|  
= a \sqrt 3 /4 \nonumber \\
J_{11}^{(0)}{\bf R},{\bf R}') &=& J_{22}^{(0)}{\bf R},{\bf R}') = J^{(0)} \ ,
\ \  {\rm if} \ |{\bf R} - {\bf R}'| = a/\sqrt 2 \ ,
\nonumber \\
J_{11}^{(0)}{\bf R},{\bf R}') &=& J_{22}^{(0)}{\bf R},{\bf R}') = M^{(0)} \ ,
\ \  {\rm if} \ |{\bf R} - {\bf R}'| = a \ .
\label{EXCHANGE} \end{eqnarray}
In other words we have exchange couplings, $K^{(0)}$, $J^{(0)}$, and $M^{(0)}$
between nn's, nnn's, and a shell of fourth-nearest neighbors (fnn's),
respectively, and these are shown in Fig. \ref{3J}.
Equation (\ref{HAM}) implies that positive exchange constants are
antiferromagnetic.  Our $J^{(0)}$ and $M^{(0)}$ correspond to YN's $-J_1$ and
$-J_2$, respectively and their simple fcc structure did not have a 
$K^{(0)}$ interaction. As will become apparent below, we include a fnn
interaction rather than a third neighbor (tnn) interaction
in the interest of algebraic simplicity.

The 4f electron of the Ce ion has quantum numbers $L=3$, $S=1/2$, and
$J=|{\bf L}+{\bf S}|=L-S=5/2$, so that ${\bf S}_{\rm op}= (g_J-1){\bf J}$,
where $g_J=6/7$ is the Land\'e $g$ factor.\cite{ASH}  The crystal field
then splits the six states of the $J=5/2$ manifold into a ground
doublet and an excited quartet state at an excitation energy of about
100K in temperature units.\cite{SPHT1,SPHT2,RES1,RES2} This
ground doublet can be described by an effective spin operator
${\bf S}_{\rm eff}$ of magnitude 1/2 and within the doublet
${\bf J}= (5/3){\bf S}_{\rm eff}$ when admixtures from the
quartet state are neglected.  In that case
\begin{eqnarray}
{\bf S}_{\rm op} = (5/3)(g_J-1){\bf S}_{\rm eff} \equiv
g_0 {\bf S}_{\rm eff} \ .
\label{gzero} \end{eqnarray}
When admixtures caused by the actual exchange field and also
an applied field of 45 kOe were calculated by Barbara {\it et al.},\cite{JS}
the moment was found to be somewhat larger than that zero net field, but
for zero applied field we neglect this effect.  Then we write the Hamiltonian
in terms of effective spins 1/2 as
\begin{eqnarray}
{\cal H} &=& \sum_{{\bf R}, n ; {\bf R}', n'} J_{n,n'}({\bf R}, {\bf R}')
{\bf S}_{\rm eff}({\bf R}+\tauv_n) \cdot {\bf S}_{\rm eff}
({\bf R}'+\tauv_{n'} ) \ ,  \label{HAM2} \end{eqnarray}
where
\begin{eqnarray}
J_{nn'}({\bf R},{\bf R}') = g_0^2 J_{nn'}^{(0)} ({\bf R},{\bf R}') \ ,
\label{gzeroEQ} \end{eqnarray} 
and we have the interactions $K$, $J$, and $M$ analogous to those
in Eq. (\ref{EXCHANGE}).

We now develop the Landau expansion for the free energy. The approach
we follow is to write the trial free energy as
\begin{eqnarray}
F &=& {\rm Tr} \left[ \rhov {\cal H} + kT \rhov \ln \rhov \right]
\end{eqnarray}
where $\rhov$ is the trial density matrix which is Hermitian and
has unit trace. The actual free energy is the minimum of $F$
with respect to the choice of $\rhov$.  Mean field theory is obtained
by restricting $\rhov$ to be the product of single-spin density
matrices, so that
\begin{eqnarray}
\rhov &=& \prod_{{\bf R},n} \rhov({\bf R},\tauv_n) \ ,
\end{eqnarray}
where $\rho({\bf R},\tauv_n)$ is the density matrix for the Ce spin
at ${\bf R}+\tauv_n$.  We write
\begin{eqnarray}
\rho({\bf R}+\tauv_n) &=& {1 \over 2} \left[ 1 + {\bf a}({\bf R}+\tauv_n)
\cdot {\bf S}({\bf R}+\tauv_n)\right] \ ,
\end{eqnarray}
where from now on ${\bf S}({\bf R}+\tauv_n)$ denotes the effective
spin 1/2 operator for the site
in question and we identify the vector trial parameter ${\bf a}$ by
relating it to the thermal expectation value of the spin as
\begin{eqnarray}
\langle {\bf S}({\bf R}+\tauv_n) \rangle &=& {\rm Tr} \left[
\rhov({\bf R}+\tauv_n) {\bf S}({\bf R}+\tauv_n) \right] \nonumber \\
&=& {\bf a}({\bf R}+\tauv_n)/4 \ , 
\end{eqnarray}
so that
\begin{eqnarray}
\rho({\bf R}+\tauv_n) &=& {1 \over 2} \left[ 1 + 4 \langle 
{\bf S}({\bf R},\tauv_n)\rangle {\bf S}({\bf R},\tauv_n)\right] \ .
\end{eqnarray}
Then, one finds that
\begin{eqnarray}
F &=& {1 \over 2} \sum_{{\bf R}, n ; {\bf R}', n'}
J_{n,n'}({\bf R}, {\bf R}') \langle {\bf S}({\bf R}+\tauv_n) \rangle
\cdot \langle {\bf S}({\bf R}'+\tauv_{n'} ) \rangle \nonumber \\
&& \ -TS \ ,
\end{eqnarray}
where 
\begin{eqnarray}
-TS &=& kT \sum_{{\bf R},n} {\rm Tr} \Biggl\{ {1 \over 2} \Biggl( 1 + 4 \langle 
{\bf S}({\bf R},\tauv_n)\rangle {\bf S}({\bf R},\tauv_n) \Biggr)
\nonumber \\ && \ \times \ln
\Biggl( {1 \over 2} \left[ 1 + 4 \langle 
{\bf S}({\bf R},\tauv_n)\rangle {\bf S}({\bf R},\tauv_n)\right] 
\Biggr) \Biggr\} \ ,
\end{eqnarray}
which we evaluate as
\begin{eqnarray}
-TS &=& kT \sum_{{\bf R},n} \sum_{p=1}^\infty
{[4 \langle S({\bf R}+\tauv_n) \rangle
\cdot \langle S({\bf R}+\tauv_n) \rangle ]^p \over 2p(2p-1)} \ .
\label{TSEQ} \end{eqnarray}
In this section we consider the term quadratic in the spin variable
and in the next section we consider the quartic term in this expansion.
(Higher order terms are not necessary for our analysis.)

\begin{figure}
\begin{center}
\includegraphics[width=7cm]{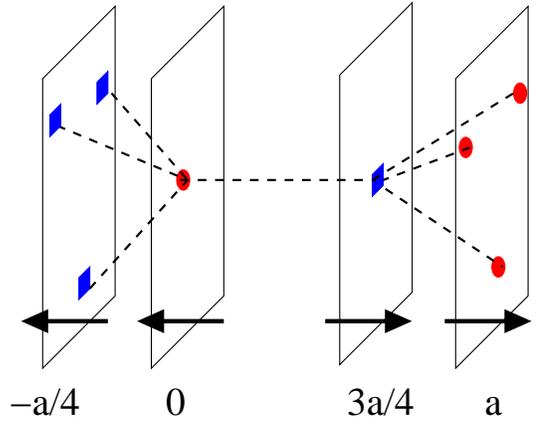} 
\caption{\label{K} (Color online) Nearest neighbor interactions $K$
(indicated by dashed lines). Here we show planes perpendicular to (111),
with Ce(1) sites represented by circles (red online) and Ce(2) sites
represented by
squares (blue online).  Below each plane the value of $x+y+z$ for that
plane is given. If the wavevector is assumed to be
$(\pi,\pi,\pi)/a$, then all spins within a given plane are parallel
to one another and if $K$ is negative (ferromagnetic), then the
spin directions of each plane are as indicated by the arrows.
Had $K$ been of opposite sign, then the spin directions of the
Ce(2) would be reversed and an inequivalent magnetic structure would
be realized.}
\end{center}
\end{figure}

We introduce as order parameters, the Fourier coefficients defined 
for $n=1,2$ by
\begin{eqnarray}
\langle {\bf S} ({\bf R} + \tauv_n) \rangle
 &=& S_n ( {\bf q}) e^{i {\bf q}\cdot {\bf R}}
+ S_n ( {\bf q})^* e^{-i {\bf q}\cdot {\bf R}} \ .
\end{eqnarray}
Note that the phase factor is determined by the origin of the unit
cell and {\it not} by the actual location of the spin site.
Any two wavevectors which differ by a linear combination of
reciprocal lattice basis vectors, ${\bf G}_n$ are equivalent, where
\begin{eqnarray}
{\bf G}_1 &=& 2 \pi (1,1,-1)/a \nonumber \\
{\bf G}_2 &=& 2 \pi (1,-1,1)/a \nonumber \\
{\bf G}_3 &=& 2 \pi (-1,1,1)/a \ .
\end{eqnarray}
Now we write the contribution to the free energy which depends on
the order parameter for some wavevector ${\bf q}$.  In terms of this order
parameter, the mean field free energy at quadratic order, $F_2$,
can be written as
\begin{eqnarray}
F_2 &=& {1 \over 2} \sum_{n,n'} [\chiv^{-1}]_{n,n'}
S_n({\bf q})^* S_{n'} ({\bf q}) \ ,
\end{eqnarray}
where\cite{FN}
\begin{eqnarray}
\chiv^{-1}({\bf q})  &=& \left[ \begin{array} {c c } 4kT + J_{11}({\bf q}) &
J_{12}({\bf q}) \\ J_{21}({\bf q}) & 4kT + J_{22}({\bf q}) \\
\end{array} \right] \ ,
\label{CHIVEQ} \end{eqnarray}
where
\begin{eqnarray}
J_{11}({\bf q}) &=& 2M (\cos q_x a + \cos q_y a + \cos q_z a) \nonumber \\
&& \ +4J [ c_x c_y + c_x c_z + c_y c_z ]  \nonumber \\ &=&
J_{22}({\bf q}) \nonumber \\
J_{12}({\bf q}) &=& K[ 1 + e^{-i(q_x+q_y)a/2} + e^{-i(q_x+q_z)a/2}
\nonumber \\ && \ + e^{-i(q_y+q_z)a/2} ] = J_{21}({\bf q})^* \ ,
\end{eqnarray}
where $c_\alpha=\cos(q_\alpha a/2)$.
Omitting the factor $4kT$, the minimum eigenvalue of the $\chiv^{-1}$
matrix, which selects the wavevector, is
\begin{eqnarray}
\lambda ({\bf q}) &=& J_{11}({\bf q}) - | J_{12}({\bf q})| \nonumber \\
&=& 2M(2c_x^2 + 2c_y^2 + 2c_z^2 - 3) \nonumber \\ && \
+ 4J(R^2 -1) - 2|K| R \ ,
\label{LAMBDAMIN} \end{eqnarray}
where
\begin{eqnarray}
R &=& \left[ 1 + c_x c_y + c_x c_z + c_y c_z \right]^{1/2} \ .
\end{eqnarray}
(By the square root, we always mean the positive square root.)
Note that changing the signs of all the $c$'s corresponds to
adding a reciprocal lattice vector to ${\bf q}$ and does not
change $\lambda ({\bf q})$.  So solutions which differ by
changing the signs of all the $c$'s are equivalent to one another.
Although the minimum value of the free energy does not depend
on the sign of $K$, the ratio of spin amplitudes within the
unit cell does depend on this sign.  To discuss the sign of $K$
it is convenient to set ${\bf q}=(1,1,1)(\pi /a)$
(which is nearly the wavevector of interest).
Then if $K$ is negative (ferromagnetic), $J_{12}({\bf q})$ is negative and the
minimal spin eigenvector is $(1,1)$, which indicates that the spins
at $\tauv_1$ and $\tauv_2$ are parallel, as is illustrated in Fig. \ref{K},
whereas if $K$ is positive,
they are antiparallel. In the former (latter) case, the other three spins of
the nn tetrahedron are antiparallel (parallel) to the spin at the origin.
Thus the sign of $K$ is easily related to whether the majority of
the nn's are parallel in which case $K$ is negative.  Otherwise
$K$ is positive.  The structure determinations indicate that
the correct choice is that $K$ is negative (ferromagnetic).
Henceforth $K$ will be used to denote $|K|$.

As mentioned in the introduction, this system for $K=0$ has
been comprehensively analyzed by YN.  However, they seem to
have overlooked an amusing limit for $K=0$.  Namely, if
$J=2M$ we have
\begin{eqnarray}
\lambda({\bf q}) &=& -3J + 2J(c_x+c_y+c_z)^2 \ .
\label{MULTIEQ} \end{eqnarray}
For $J<0$ this is minimal for $c_x=c_y=c_z=\pm 1$.  For
$J>0$ this is minimized over the entire two dimensional
manifold for which $c_x+c_y+c_z=0$.  What this means
is that for this special case, there is no wavevector
selection.  Such a multiphase point has been found in 
several models.\cite{KAG,PYR,NOQ}
As we shall see, this multiphase behavior is modified to
encompass a one dimensional manifold when $K$ is small
and $M< J/2$.

\subsection{$J<0$}

When $J<0$, then the second and third terms of Eq. (\ref{LAMBDAMIN})
do not compete with one another:  for a fixed value of $c_x^2+c_y^2+c_z^2$,
$\lambda({\bf q})$ is minimized by maximizing $R$, which implies that
$c_x=c_y=c_z\equiv c$, so that
\begin{eqnarray}
\lambda({\bf q}) = -6M + 12(J+M)c^2 -2K\sqrt{1+3c^2} \ .
\end{eqnarray}
The extrema must be either for $c^2=0$, $c^2=1$, or (by differentiation)
\begin{eqnarray}
3c^2 &=& -1 + \left[ {K \over 4(J+M)} \right]^2 \ ,
\end{eqnarray}
so that, for this to apply, we must satisfy
\begin{eqnarray}
4(J+M) < K < 8(J+M) \ .
\end{eqnarray}
To represent the results it is convenient to set the magnitude of
$J$ equal to unity, or, here, $J=-1$. In this case $M$ is restricted by
\begin{eqnarray}
1+ K/8< M < 1 + K/4 \ ,
\end{eqnarray}
in which case this value of $c$ gives

\begin{center}
\begin{figure} [h]
\includegraphics[scale=0.7]{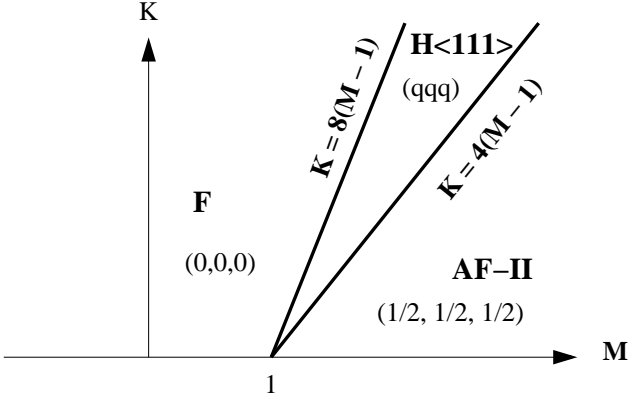}
\caption{\label{MINM} Minimum free energy configurations for $J=-1$ as
a function of $M$ and $K$. The wavevectors $\hat {\bf q}$
are indicated along with the labeling of YN for the phases.}
\vspace{0.02 in}
\end{figure}
\end{center}

\begin{eqnarray}
\lambda({\bf q}) &=& -10 M -4 - {K^2 \over 4(M-1)} \ .
\end{eqnarray}
When we compare this result with the value of $\lambda ({\bf q})$
which we get for ${\bf q}_\alpha =0$ and for ${\bf q}_\alpha = \pi /2$,
we get the phase diagram shown in Fig. \ref{MINM}.

\subsection{$J>0$}

For positive $J$ the minimization is more complicated because the
second and third terms in Eq. (\ref{LAMBDAMIN}) now compete.
For $\lambda({\bf q})$ to be extremal, its gradient with respect to
${\bf q}$ must vanish, so that
\begin{eqnarray}
0 &=& s_x\Biggl[ -8Mc_x -4J(c_y+c_z) +  (c_y+c_z)K/R \Biggr] \ , \nonumber \\
0 &=& s_y \Biggl[ -8Mc_y -4J(c_x+c_z) +  (c_x+c_z)K/R \Biggr] \ , \nonumber \\
0 &=& s_z \Biggl[ -8Mc_z -4J(c_x+c_y) +  (c_x+c_y)K/R \Biggr] \  .
\end{eqnarray}

There are obviously many subcases for the extrema and we will not
consider equivalent solutions which correspond to changing the
signs of all the $c_\alpha$'s or permuting their subscripts.
Thus we have four cases:
\begin{eqnarray}
s_x=s_y=s_z=0\ , \ \ \ \ {\rm Case \ \ I} \ ,
\end{eqnarray}
\begin{eqnarray}
s_x=s_y=0 \ , \ s_z \not=0 \ , \ \ \ \ {\rm Case \ \ II} \ , 
\end{eqnarray}
\begin{eqnarray}
s_x=0 \ ,  \ \ \ \ s_y \not= 0 \ , \ \ \ s_z\not=0\ , \ \ \ \ {\rm Case \ \ 
III} \ ,
\end{eqnarray}
and
\begin{eqnarray}
s_x\not= 0 \ , \ \ \  s_y\not=0 \ , \ \ \ \
s_z\not= 0\ , \ \ \ \ {\rm Case \ \ IV} \  .
\end{eqnarray}

\subsubsection{Case I}

When $s_x=s_y=s_z=0$, then $c_x$, $c_y$, and $c_z$ can each assume
the values $+1$ and $-1$.  So we have
\begin{eqnarray}
c_x&=&c_y=c_z = 1 \ , \ \ \ \ \ {\rm Case \ \ Ia} \ , \nonumber \\
c_x&=&c_y=-c_z = 1 \ , \ \ \ \ \ {\rm Case \ \ Ib} \ ,
\end{eqnarray}
so that
\begin{eqnarray}
\lambda &=& 6M + 12 J - 4K \ , \ \ \ \ \ {\bf q}=(0,0,0) \ , \nonumber \\
\lambda &=& 6M -4J \ , \ \ \ \ \ {\bf q}=(0,0,2 \pi /a) \ .
\end{eqnarray}
Case Ia, ${\bf q}=(0,0,0)$, is the F phase of YN and Case Ib,
${\bf q}=(2 \pi/a,0,0)$, the AF-I phase of YN.

\subsubsection{Case II}

In this case, $q_x$ and $q_y$ can independently assume the values
$0$ or $2 \pi /a$, so that $c_x$ and $c_y$ independently assume
the values $+1$ and $-1$.  Then we have 
\begin{eqnarray}
c_x=c_y = 1 \ , \ \ \ \ {\rm Case \ \ IIa}
\end{eqnarray}
and
\begin{eqnarray}
c_x=-c_y = 1 \ , \ \ \ \ {\rm Case \ \ IIb} \ .
\end{eqnarray}

In Case IIa we have
\begin{eqnarray}
\lambda ({\bf q}) &=& 2M (1+2c_z^2) + 4J(1 +2 c_xc_z)
\nonumber \\ && \ -2K \sqrt{ 2 + 2 c_z c_x} \ ,
\end{eqnarray}
so that minimization with respect to $q_z$ yields
\begin{eqnarray}
0 &=& \Biggl( -8Mc_z  - 8J c_x + {2Kc_x \over
\sqrt{2 + 2 c_z c_x} } \Biggr)  s_z =0 \ .
\end{eqnarray}
So either $s_z=0$ (which repeats Case I), or (since $c_x^2=1$)
\begin{eqnarray}
8Mc_zc_x + 8J &=& {2K \over \sqrt{2 + 2 c_zc_x}} \ .
\end{eqnarray}
This gives
\begin{eqnarray}
32 M^2 c_z^2 + 64 JMc_xc_z + 32{J}^2 &=& {K^2 \over 1 + c_zc_x} \ .
\label{IIa} \end{eqnarray}
For $K=0$, this gives $c_xc_z=-J/M$ and
\begin{eqnarray}
\lambda ({\bf q}) &=& 2M + 4J - {4J}^2/M \ ,
\end{eqnarray}
which is the H$<100>$ phase of YN with wavevector $(q_x,0,0)$.
For $K=0$ this has no range of stability.  For $K \not=0$ we
evaluate $\lambda({\bf q})$ by solving Eq. (\ref{IIa}) numerically.

In Case IIb we have
\begin{eqnarray}
\lambda({\bf q}) &=& 2M (1 + 2 c_z^2) -4J \ .
\end{eqnarray}
For positive $M$ we thus have ${\bf q}=(0,2,1)\pi/a$, which
is the AF-III phase of YN and
\begin{eqnarray}
\lambda ({\bf q}) &=& 2M-4J \ , \ \ \ \ {\rm Case \ \ IIb} \ .
\end{eqnarray}
We discard the case when $M$ is negative because it repeats case Ib.

\subsubsection{Case III}

Here $c_x=\pm 1$ (we need only consider $c_x=+1$)
and $c_y$ and $c_z$ are nonzero, determined by
\begin{eqnarray}
0 &=& -8Mc_y -4J(c_x+c_z) +  (c_x+c_z)K/R \ , \nonumber \\
0 &=& -8Mc_z -4J(c_x+c_y) +  (c_x+c_y)K/R \  .
\end{eqnarray}
Subtracting and adding one equation from the other we get
\begin{eqnarray}
8M(c_y-c_z) &=& 4J(c_y-c_z) - K(c_y-c_z)/R \nonumber \\
8M(c_y+c_z) &=& (2c_x+ c_y+c_z)(-4J+2K/R) \ ,
\end{eqnarray}
so that
\begin{eqnarray}
c_x &=& 1 \ , \ \ \ c_y=c_z \ , \ \ \ \ \ {\rm Case \ \ IIIa}
\nonumber \\ 8M(c_y+c_z) &=& (2+c_y + c_z)(-4J + K/R) \ , \\
c_x &=& 1 \ , \ \ \ 8M=4J- K/R \ , \ \ \ \ \ {\rm Case \ \ IIIb}
\nonumber \\ 8M(c_y+c_z) &=& (2+c_y + c_z)(-4J + K/R) \ .
\label{EQIIIb} \end{eqnarray}

In Case IIIa we have $c_x=1$, $c_y=c_z$, where
\begin{eqnarray}
8M c_y &=& -4J(1+c_y) + K \ ,
\end{eqnarray}
so that
\begin{eqnarray}
c_z= c_y = (K- 4J)/(8M+4J) \ , \ \ \ \ {\rm Case \ \ IIIa} \ .
\end{eqnarray}
For this case to apply, we must satisfy the restriction
\begin{eqnarray}
|K- 4J| < |8M+4J| \ .
\end{eqnarray}
Then we obtain
\begin{eqnarray}
\lambda ({\bf q}) &=& -2M -2K - {(K-4J)^2 \over 8M+4J} \ .
\end{eqnarray}
For $K=0$ this solution is H$<110>$ of YN.

In Case IIIb we have $c_x=1$,
\begin{eqnarray}
8M=4J - K/ [ (1+c_y)(1+c_z)]^{1/2} \ ,
\label{EQAA} \end{eqnarray}
and Eq. (\ref{EQIIIb}) becomes
\begin{eqnarray}
8M(c_y+c_z) &=& -8M(2+c_y+c_z) \ . 
\end{eqnarray}
This gives $M=0$ or
\begin{eqnarray}
c_y+c_z &=& -1 \ .
\end{eqnarray}
Since $c_x+c_y+c_z=0$, which will appear as Case IVc, below, we do not
consider it further here.

\subsubsection{Case IV}

Here
\begin{eqnarray}
0 &=& -8Mc_x -4J(c_y+c_z) +  (c_y+c_z)K/R \ , \nonumber \\
0 &=& -8Mc_y -4J(c_x+c_z) +  (c_x+c_z)K/R \ , \nonumber \\
0 &=& -8Mc_z -4J(c_x+c_y) +  (c_x+c_y)K/R \  .
\label{EQAD} \end{eqnarray}
This set of equations is of the form
\begin{eqnarray}
\left( \begin{array} {c c c} A & B & B \\ B & A & B \\
B & B & A \\ \end{array} \right) \left[ \begin{array} {c}
c_x \\ c_y \\ c_z \\ \end{array} \right] = 0 \ ,
\label{TYPE} \end{eqnarray}
where $A= -8M$ and $B= -4J + K/R$.  Note that the eigenvalues of this
matrix are $A+2B$, $A-B$, and $A-B$.  The solution of this set of
equations is either of type a (in which $c_x=c_y=c_z=0$),
type b (in which the eigenvalue $A+2B$ is zero), or type c,
(in which the eigenvalue $A-B$ is zero).

The solution of type a is case IVa, with
\begin{eqnarray}
c_x=c_y=c_z = 0 \ , \ \ \ \ \lambda({\bf q}) = - 6M -2K \ .
\end{eqnarray}
For $K=0$, this is AF-II of YN.

The solution to Eq. (\ref{TYPE}) of type b is Case IVb with
\begin{eqnarray}
(c_x, c_y, c_z) &=& (c,c,c) \ , \ \ \ \ A+2B = 0 \ .
\end{eqnarray}
Setting $A+2B=0$ leads to
\begin{eqnarray}
c &=& \left[ {1 \over 3} \right]^{1/2}
\left[ \left( {K \over 4(J+M)} \right)^2 -1 \right]^{1/2} \ ,
\end{eqnarray}
and we have the constraint
\begin{eqnarray}
1 < {K \over 4(J+M)} < 2 \ ,
\end{eqnarray}
and therefore this regime does not appear in the limit studied by YN.
Then
\begin{eqnarray}
\lambda({\bf q}) &=& -10M -4J - {K^2 \over 4(J+M)} \ .
\end{eqnarray}

The solution to Eq. (\ref{TYPE}) of type c requires $A-B=0$ and
the solution must be a linear combination of the two
associated eigenvectors.  So we introduce Potts-like variables\cite{POTTS}
\begin{eqnarray}
(c_x, c_y, c_z) &=& \left( -{\alpha \over \sqrt 2} - {\beta \over \sqrt 6} \ ,
{\alpha \over \sqrt 2} - {\beta \over \sqrt 6} \ , {2 \beta \over \sqrt 6}
\right) \ .
\label{CASEIVc} \end{eqnarray}
Note that we have
\begin{eqnarray}
c_x^2 + c_y^2 + c_z^2 &=& \alpha^2 + \beta^2
\end{eqnarray}
and
\begin{eqnarray}
c_x c_y + (c_x+c_y)c_z &=& - {1 \over 2} (\alpha^2 + \beta^2 ) \ .
\end{eqnarray}
The equation $A-B=0$ is
\begin{eqnarray}
0 &=& 4J- 8M - {K \over \sqrt{1 -(\alpha^2 + \beta^2)/2}} \ .
\label{MINIM} \end{eqnarray}
If we write
\begin{eqnarray}
4J=8M+\xi K \ ,
\label{XIDEF} \end{eqnarray}
where $\xi$ can not be negative, then
\begin{eqnarray}
\alpha^2 + \beta^2 = 2 - 2 \xi^{-2} \equiv X^2 \ .
\label{EQXI} \end{eqnarray}

\noindent
This indicates that $\xi > 1$ or
\begin{eqnarray}
0 < {K \over 4J-8M} < 1 \ .
\label{CONSTEQ} \end{eqnarray}
Thus we set
\begin{eqnarray}
\alpha &=& X \cos \theta \ , \ \ \ \ \beta = X \sin \theta \ ,
\end{eqnarray}
where the restriction on $\theta$ will be discussed.  These evaluations give
\begin{eqnarray}
X^2 = 2 - 2 \left[ {K \over 4J-8M} \right]^2 
\equiv 2-2(K/K_c)^2 \ ,
\end{eqnarray}
so that
\begin{eqnarray}
\lambda({\bf q})
&=& 2M- 4J - {K^2 \over 4J-8M } \  , \ \ \ \ {\rm Case \ \ IVc} 
\end{eqnarray}
and we have the constraint of Eq. (\ref{CONSTEQ}).  Then, Eq. 
(\ref{CASEIVc}) gives
\begin{eqnarray}
{\bf c} &\equiv& (c_x,c_y,c_z) \nonumber \\ &=& {2X \over \sqrt 6}
[ \sin \left( \theta - 2 \pi /3 ) , \sin (\theta +2 \pi /3)
, \sin \theta  \right] \ .
\label{DKEQ} \end{eqnarray}

Note that $\theta$ is arbitrary, so this minimum is realized along a curve
in wavevector space.  Equation (\ref{EQXI}) shows that $X \leq \sqrt 2$. 
As long as $X$ is less than $\sqrt 6/2$ ({\it i. e.} $K_c>K>K_c/2$), all
values of $\theta$ are acceptable.  If $X$ lies between $\sqrt 6 /2$
and $\sqrt 2$ ({\it i. e.} $0<K<K_c$), then values of $\theta$
symmetric around $\theta=0$ (and also around
$\theta = k\pi/3$,

\begin{center}
\begin{figure} [h]
\includegraphics[scale=0.60]{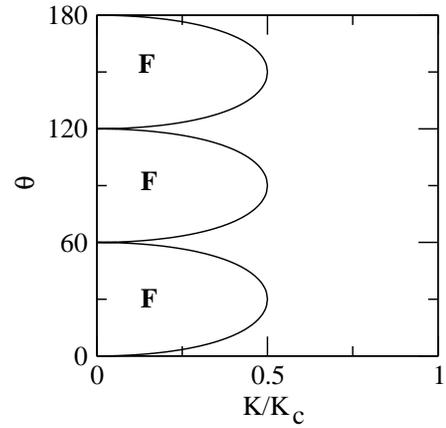}
\caption{\label{CIRCLE} Allowed and forbidden (indicated by an F)
regions $\theta$ (in degrees) versus $K/K_c$, where $K_c=4J-8M$.}
\vspace{0.02 in}
\end{figure}
\end{center}

\noindent
where $k$ is an integer) in which none of the $c$'s exceed one in magnitude
are allowed. The allowed region, $-\theta_c < \theta < \theta_c$, is
determined by the condition
\begin{eqnarray}
(2 / \sqrt 6) X \sin(\theta_c + \pi/3) = 1 \ .
\end{eqnarray}
The allowed regions of $\theta$ are illustrated in the Fig.  \ref{CIRCLE}.
We will call this phase the M' phase.

\begin{table*}
\begin{center}
\caption{\label{RESULTS} Free energies of the various states.}
\begin{tabular} {|| c || c | c | c || c ||}
\hline \hline
Case & Energy & $\hat {\bf q}$ & For & \ \ YN Case$^a$\ \  \\ \hline
Ia & $6M+12J-4K$ & $(0,0,0)$ & Any & F \\ 
Ib & $6M -4J$ & $(1,0,0)$ & Any & AF-I \\
IIa & Complex$^b$ & $(q_x,0,0)$ & $|J/M|<1^{\rm c}$ & H$<$100$>$ \\
IIb & $2M-4J$ & (0,1,1/2) & Any & AF-III \\
IIIa & \ \ $-2M-2K- {(K-4J)^2 \over 8M+4J}$\ \ & $(0,q_y,q_y)$ & 
\ \ $\left| {K-4J) \over (8M+4J)} \right| < 1$ \ \ & H$<$110$>$ \\
IVa & $-6M-2K$ & $(1/2,1/2,1/2)$ & Any & AF-II \\
IVb & $-10M -4J - {K^2 \over 4(J+M)}$ & 
$(q_x,q_x,q_x)$ & ${1 \over 2} < {4(J+M) \over K} < 1 $ & \\
IVc & $2M-4J - {K^2 \over 4J-8M}$ & \ \ $\left( {1\over 2} -\delta,
{1\over 2} +\delta, {1\over 2} \right)^{\rm d}$\ \ &
$K < 4J-8M$ & \\
\hline
\end{tabular}
\end{center}

\vspace{0.2 in}
\noindent
a For $K=0$.

\noindent
b For $K=0$, the energy is $4J+2M-4J^2/M$.

\noindent
c This restriction is only for $K=0$.

\noindent
d For $K=0$ we have a two-parameter multiphase: $c_x+c_y+c_z=0$.  For
$K \not= 0$, the single parameter $\theta$ of Eq. (\protect{\ref{DKEQ}})
is not fixed.
\end{table*}

Although $\theta$ is arbitrary, to get the
wavevector seen by experiment, we want to have 
\begin{eqnarray}
{\bf q}a = (\pi - 2\pi \delta  \ , \pi + 2\pi \delta , \pi ) \ , 
\end{eqnarray}
so that
\begin{eqnarray}
c_x &=& \cos (\pi /2 - \pi \delta) =  \sin (\pi \delta ) \ , \nonumber \\
c_y &=& \cos (\pi /2 + \pi \delta) = - \sin (\pi \delta ) \ , \nonumber \\
c_z &=& \cos (\pi /2 ) = 0 \ .
\end{eqnarray}
This corresponds to $\theta=\pi $ in Eq. (\ref{DKEQ}), so that
\begin{eqnarray}
\sin (\pi \delta ) &=& (2X/ \sqrt 6) \sin (\pi /3) = (X/ \sqrt 2) 
\nonumber \\ &=& \left[ 1 - \left( {K \over 4J-8M} \right)^2 \right]^{1/2} 
=  \left[ 1 - \left( {K \over K_c} \right)^2 \right]^{1/2} \ .
\end{eqnarray}
Presumably fluctuation effects or further neighbor interactions
select $\theta= \pi $ from the degenerate
manifold of all $\theta$ which minimize $\lambda({\bf q})$
and we will consider the second mechanism in the Sec. III.

\subsubsection{Comparison of Extrema}

To find the global minimum of the eigenvalue, we must
compare the values of the functions at the above extrema.
For this purpose we summarize the results in Table \ref{RESULTS}.

\begin{center}
\begin{figure} [ht]
\includegraphics[scale=0.5]{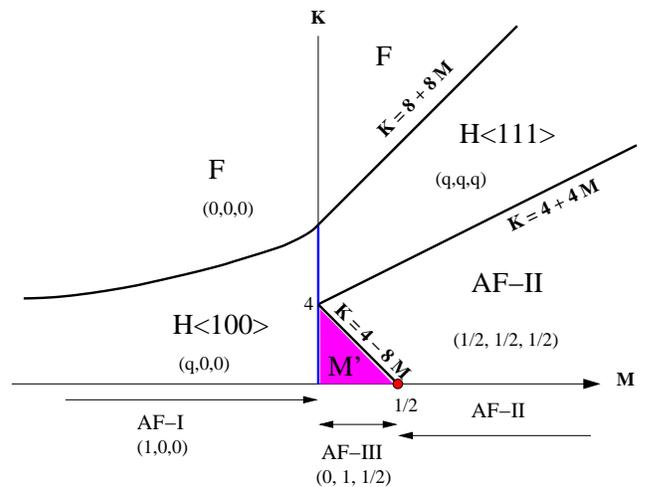}
\caption{\label{MINP} (Color online) Minimum free energy configurations
for $J=1$, as a function of $M$ and $K$.  The components of the wavevector 
(in units of $2 \pi /a$)  are the triad of numbers in parentheses.  The
point $K=0$, $M=1/2$ (filled circle, online red) is a
two-parameter multiphase point where all states satisfying $c_x+c_y+c_z=0$
have minimal free energy. For $K \not= 0$, the region (online magenta)
labeled $M'$ is the multiphase region in which the single parameter
$\theta$ can be chosen as in Fig. \protect{\ref{CIRCLE}}.  The line
segment (online blue) $M=0$, $K<8$ is a multiphase region
in which only the value of $R=K/4$ [see Eq. (\protect{\ref{LAMBDAMIN}})
is fixed.  The phase transitions are
continuous except for those at $M=0$. As soon as $K$ is nonzero,
the AF-I phase gives way to the H$<110>$ phase having small $q$.
and the AF-III phase gives way to the M' phase which has wavevector
within the allowed range of $\theta$ of Eq. (\protect{\ref{DKEQ}}).}
\vspace{0.02 in}
\end{figure}
\end{center}

Since Case IIa is hard to analyze analytically, we had recourse to a
computer program to compare the various local extrema and select
the global minimum.  Having done that, we checked some of the results
analytically with the result shown in Fig. \ref{MINP}.  It is
interesting that turning on $K$ immediately
renders the AF-I and AF-III phase unstable.  Note also that the M' phase
(which is the one we want for CEAL) does occur for $K$ as large as $4J$ and
$M$ being quite small.  This agrees with the idea that the interactions
decrease with increasing separation so that $|K|> |J|>M$. 

Finally, we remark that the model with only $K$ and $J$ nonzero
lacks wavevector selection for $K<8J$ because $\lambda({\bf q})$ only
depends on $R$. So this means that $c_x$, $c_y$, and $c_z$ range
over a two parameter manifold of fixed $R$.

\section{FURTHER NEIGHBOR INTERACTIONS}

In this section we consider the effect of third- and further-than-fourth-
neighbor interactions.

\subsection{A ``4$J$" MODEL}

We start by including third-neighbor (tnn)
interaction coefficients $L$, as shown in Fig. \ref{3J}, so that we
have a $4J$ model.  This interaction occurs at separations equivalent
to $(3,1,\overline 1)a/4$.  Including them does not affect
$J_{11}({\bf q})$ or $J_{22}({\bf q})$ but now
\begin{eqnarray}
J_{12}({\bf q}) &=& J_{21}({\bf q})^* =
e^{-ia(k_x+k_y+k_z)/4} [ K \Phi + L \Psi ]
\end{eqnarray}
where $\Phi$ is as before, but now
\begin{eqnarray}
\Psi &=& e(3 1 \overline 1 ) + e(3 \overline 1 1) +
e(\overline 3 1 1 ) + e(\overline 3 \overline 1 \overline 1) + {\cal P} \ ,
\end{eqnarray}
where $e(lmn) = \exp[i(lk_x+mk_y+nk_z)a/4]$ and ${\cal P}$ indicates
inclusion of the permutations ${\cal P} =
c_x \rightarrow c_y \rightarrow c_z \rightarrow c_x$.  Then
\begin{eqnarray}
\lambda({\bf k}) &=& 4J(c_xc_y+c_yc_z+c_zc_x) -6M \nonumber \\ && \
+4M(c_x^2+c_y^2+c_z^2) -2KR \ ,
\end{eqnarray}
where now
\begin{eqnarray}
R^2 &=& [1 + A ] + (L/K) \Delta + (L/K)^2 \Pi \ ,
\end{eqnarray}
with
\begin{eqnarray}
\Delta &=& {1 \over 4} \left[ \Phi^* \Psi + \Phi \Psi^* \right] =
 -6 + 2A + 4 B+ 4C 
\end{eqnarray}
\begin{eqnarray}
\Pi &=& {1 \over 4} \Psi^* \Psi \nonumber \\ &=& 9 -7A -8B +8C +4D +8E -4F \ .
\end{eqnarray}
Also
\begin{eqnarray}
A &=& c_xc_y + c_xc_z + c_yc_z \nonumber \\
B &=& c_x^2 + c_y^2 + c_z^2 \nonumber \\
C &=& (c_x+c_y+c_z) c_xc_yc_z \nonumber \\
D&=& (c_x+c_y+c_z)(c_x^3+c_y^3+c_z^3) \nonumber \\
E&=& (c_x^2c_y^2+c_y^2c_z^2+c_z^2c_x^2) \nonumber \\
F&=& (c_x^4+c_y^4+c_z^4) \ .
\end{eqnarray}

We will not pursue the analysis to the same level as for the 3J model.
Here we will show that for an interior point in $c_x,c_y,c_z$ space
(i. e. when all these variables are less than 1 in absolute value), there
is no extremum of $\lambda({\bf q})$ for which all the $c$'s are
different from one another. (Thus this model can not give the state
of the type ${\bf c}=(c,-c,0)$, observed for CEAL.\cite{CEAL5,CEAL6})  
For this analysis we consider the equations
$\partial \lambda({\bf q}) / \partial c_\alpha =0$ {\it under the assumption
that all the $c$'s are different from one another}. For
$\alpha=x$ this derivative condition is 
\begin{eqnarray}
0 &=& 4J(c_y+c_z) + 8Mc_x - {1 \over R} \Biggl[ K (c_y+c_z) \nonumber \\
&+& L(2c_x+2c_y+8c_x + 4c_yc_z(c_x+c_y+c_z) + 4c_xc_yc_z \Biggr]
\nonumber \\ &+& (L^2/K) \Biggl[ -16c_x -7c_y-7c_y -56c_x^3 +16 c_xc_y^2
+16 c_xc_z^2 \nonumber \\ &+& (12c_x^2+8c_yc_z)(c_x+c_y+c_z)
+ 4 c_x^3 + 8c_xc_yc_z \Biggr] \ .
\end{eqnarray}
We now subtract from this equation that which one gets by the permutation
${\cal P}$ and divide the result by $c_x-c_y$, a quantity which, by
assumption, is nonzero. Thereby we obtain
\begin{eqnarray}
0 &=& -4J + 8M - {1 \over R} \Biggl(
-K +L[ 6 -4c_z(c_x+c_y+c_z) ]
\nonumber \\ && \  + (L^2/K) [ -9 -56(c_x^2+c_xc_y+c_y^2) -16 c_xc_y
\nonumber \\ && \ + 16 c_z^2 + 12(c_x+c_y)(c_x+c_y+c_z) 
\nonumber \\ && -8 c_z(c_x+c_y+c_z) + 4 (c_x^2+c_y^2+c_z^2) \Biggr) \ .
\end{eqnarray}
Now, again subtract from this equation that which one gets by the permutation
${\cal P}$ and divide the result by $c_x-c_z$, a quantity which, by
assumption, is nonzero. Thereby we get
\begin{eqnarray}
0 &=& L(c_x+c_y+c_z) - 48  (L^2/K) (c_x+c_y+c_z) 
\label{CXCYCZ} \end{eqnarray}
which indicates that $c_x+c_y+c_z=0$.  Therefore we introduce the Potts
representation in the form
\begin{eqnarray}
c_x = {\xi \over \sqrt 3} + {X [\sin \theta + \sqrt 3 \cos \theta ] \over 
\sqrt 6 } \nonumber \\
c_y = {\xi \over \sqrt 3} + {X [\sin \theta - \sqrt 3 \cos \theta ] \over 
\sqrt 6 } \nonumber \\
c_z = {\xi \over \sqrt 3} - {2X \sin \theta \over \sqrt 6 } \ .
\label{EQPOTTS} \end{eqnarray}
If there is an extremum for which all the $c$'s are different from
one another, the calculation we have just done shows that it must occur for
$\xi=0$.  Rather than further analyze the derivative conditions, it
is instructive to consider $\lambda({\bf q})$ in terms of these
Potts variables.  We have
\begin{eqnarray}
A &=& \xi^2 - {1 \over2 } X^2 \ , \nonumber \\
B &=& \xi^2 + X^2 \ , \nonumber \\
C &=& {\xi^4 \over 3} - {1 \over 2} \xi^2 X^2 -
{1 \over 3 \sqrt 2} \xi X^3 \sin (3 \theta ) \ , \nonumber \\
D &=& \xi^4 +3 \xi^2 X^2 
- {1 \over \sqrt 2} \xi X^3 \sin (3 \theta ) \ , \nonumber \\
E &=& {\xi^4 \over 3} + {1 \over 4} X^4
+ {\sqrt 2  \over 3 } \xi X^3 \sin (3 \theta ) \ , \nonumber \\
F &=& {\xi^4 \over 3} + 2 \xi^2 X^2 + {1 \over 2} X^4
- {2 \sqrt 2 \over 3 } \xi X^3 \sin (3 \theta ) \ .
\end{eqnarray}
Thus
\begin{eqnarray}
\Delta &=& -6 + 6 \xi^2 + 3X^2 +
{4 \over 3} \xi^4 - 2 \xi^2 X^2 - {2\sqrt 2 \over 3}
\xi X^3 \sin (3 \theta) \ , \nonumber \\
\Pi &=& 9 -15 \xi^2 - {9 \over 2} X^2
+ 8 \xi^4 + 2 \sqrt 2 \xi X^3 \sin (3 \theta) \ .
\end{eqnarray}
The important point is that, correct to leading order in $L/K$ we have
\begin{eqnarray}
R = f(\xi^2, X^2) - {\sqrt 2 \over 3} (L/K) \xi X^3 \sin (3 \theta) \  ,
\end{eqnarray}
and therefore a contribution to $\lambda({\bf q})$ of
\begin{eqnarray}
\delta \lambda({\bf q}) \sim {2 \sqrt 2 \over 3} L \xi X^3 \sin (3 \theta ) \ .
\end{eqnarray}
Thus the fact that $L$ is nonzero leads to a nonzero term in
$\lambda({\bf q})$ which is linear in $\xi$ and renders the
manifold $c_x+c_y+c_z=0$ unstable. 
Since the quadratic term in $\xi$ is of the form 
\begin{eqnarray}
\lambda({\bf q}) \sim 1/2 \chi_\xi^{-1} \xi^2  
\end{eqnarray}
with $\chi_\epsilon^{-1} \approx 8J-2K$, we see that, when a minimization with
respect to $\xi$ is performed, one has
\begin{eqnarray}
\xi &=& - {2 \sqrt 2 \chi_\xi \over 3} L X^3 \sin (3 \theta) 
\end{eqnarray}
and now we generate the following term in $\lambda({\bf q})$
which depends on $\theta$:
\begin{eqnarray}
\lambda({\bf q}) &=& f(X^2) - {4 \chi_\xi \over 9}
L^2 X^6 \sin^2 (3 \theta) \ .
\label{THETAEQ} \end{eqnarray}
When $L$ is nonzero, it generates a nonzero value of
$\xi$, {\it i. e.} it would take the extremum slightly out of the
plane $c_x+c_y+c_z=0$.  But, according to Eq. (\ref{CXCYCZ}), the
minimum with the $c_\alpha$'s being unequal, can only occur in
the plane $c_x+c_y+c_z=0$.  So, if a minimum can not occur in
this plane, it can not occur anywhere in the interior of ${\bf c}$ space.
In addition, even if a small displacement out of this plane were
allowed (and it would not be totally unacceptable in view
of the experimental data if $\xi$ were small enough), the
$\theta$-dependent term in Eq. (\ref{THETAEQ})
favors $\theta=\pi /2$, which would give a wavevector of the form
$a{\bf q}/(2\pi ) = (1/2 - \delta, 1/2 - \delta, 1/2 +2 \delta)$, which
the experimental data do not permit. Within the 4J model
this problem can not overcome because the sign of this anisotropy in
$\theta$ can not be adjusted (it enters in terms of positive
definite quantities).

\subsection{STILL FURTHER NEIGHBOR INTERACTIONS}

The preceding calculation, although unsuccessful in producing an
explanation of the data, is nevertheless instructive.
It indicates that we need to focus on the sixth order anisotropy
(in {\bf c} space)
coming from further neighbor interactions.  This requires a term involving
six powers of the $c$'s.  The leading candidate for such a term is
the exchange interaction at the separation $(a,a,a)$.  From
these eight equivalent neighbors, with exchange constant $Q$,
one finds the additional contribution to
$J_{nn}({\bf q})$ to be
\begin{eqnarray}
\delta J_{nn}({\bf q}) &=& 8 Q \cos (aq_x) \cos (aq_y) \cos (aq_z)
\nonumber \\ &=& 8Q(2 c_x^2-1)(2c_y^2-1)(2c_z^2-1) \ ,
\label{QEQ} \end{eqnarray}
which leads to an additional term in $\lambda({\bf q})$ whose
dependence on $\theta$ is of the form
\begin{eqnarray}
\delta \lambda({\bf q}) &=& 64 Q c_x^2 c_y^2 c_z^2 \nonumber \\
&=& Q \Biggl( {8 \xi^3 \over 3 \sqrt 3} - {8\xi X^2 \over 2 \sqrt 3}
- {8 \over 3\sqrt 6} X^3 \sin (3 \theta) \Biggr)^2 \nonumber \\
&\sim& {32 \over 27} Q X^6 \sin^2 (3 \theta) \ .
\label{ANISOTROPY} \end{eqnarray}
(Here we dropped the less significant terms proportional to $\sin(3 \theta)$.)
The sign of this term {\it is} adjustable and it will have the
opposite sign from the anisotropy due to $L$ if $Q$ is positive
(antiferromagnetic).  It will then favor $\theta = n\pi/3$
in Eq. (\ref{EQPOTTS}) and if this anisotropy dominates,
then the wavevector will be of the desired form:
$a{\bf q}/(2 \pi) = (1/2-\delta, 1/2+\delta, 1/2)$.

If we only consider $L$ and $Q$, then the condition that
this anisotropy have the correct sign to explain the wavevectors of
CEAL is that
\begin{eqnarray}
Q  > (3/8) \chi_\xi L^2 \ .
\label{BOUND} \end{eqnarray}
Since the model now has four parameters, we did not pursue a definitive
numerical analysis of the minima.  However, to corroborate this
argument is sound, we give in Table \ref{EXTABLE} some values of the
input parameters which give ${\bf c}=(c,-c,0)$, for $c$ equal to
the experimental value $c= \sin \pi \delta= 0.338$ for
$\delta = 0.11$.\cite{CEAL5,CEAL6} This table illustrates the
phase transition in the anisotropy in ${\bf c}$-space which
takes place for small $Q$ at a value close to that predicted
by the approximate bound of Eq. (\ref{BOUND}).  Note that for small
values of $L$ and $Q$, the values of the other parameters which give
the desired form of ${\bf c}$ can not be far from the region M'
of Fig. \ref{MINP}.

\begin{table}
\caption{\label{EXTABLE} Values of the exchange integrals (for $J=1$)
which give values of ${\bf c}$ close to the observed 
values of ${\bf c}$.  Note that a small change in $Q$ causes the
anisotropy in ${\bf c}$-space to change.}
\vspace{0.2 in}
\begin{tabular} {|| c c c c | c c c ||} \hline \hline
 $|K|$  & $L$ & $M$  & $Q$  & $c_x$  & $c_y$ & $c_z$ \\ \hline
 $3.50^{\rm a}$ &$ -0.04 $&$  0.016 $&$  0.0010 $&$ -0.339$&$  0.001$&$  0.338$\\
 3.50 &$ -0.04 $&$  0.016 $&$ -0.0010 $&$ -0.425$&$  0.215$&$  0.215$\\ 
 3.50 &$ -0.04 $&$  0.017 $&$  0.0010 $&$ -0.333$&$  0.000$&$  0.333$\\
 3.50 &$ -0.03 $&$  0.016 $&$  0.0010 $&$ -0.360$&$  0.000$&$  0.360$\\
\hline
 $3.50^{\rm a}$ &$ -0.08 $&$ -0.012 $&$  0.0050 $&$ -0.337$&$  0.000$&$  0.337$\\
 \hline 3.00 &$  0.08 $&$  0.131 $&$  0.0010 $&$ -0.344$&$  0.022$&$  0.321$\\
 3.00 &$  0.08 $&$  0.131 $&$  0.0006 $&$ -0.358$&$  0.033$&$  0.324$\\
 3.00 &$  0.08 $&$  0.131 $&$  0.0000 $&$ -0.410$&$  0.203$&$  0.206$\\ \hline
 $3.00^{\rm a}$ &$  0.00 $&$  0.102 $&$  0.0001 $&$ -0.333$&$  0.000$&$  0.333$\\
\hline 
$3.00^{\rm a}$ &$  0.04 $&$  0.116 $&$  0.0004 $&$ -0.340$&$  0.000$&$  0.340$\\ \hline
$3.00^{\rm a}$ &$ -0.04 $&$  0.085 $&$  0.0003 $&$ -0.336$&$  0.000$&$  0.336$\\
\hline
$3.00^{\rm a}$ &$ -0.08 $&$  0.066 $&$  0.0012 $&$ -0.338$&$  0.000$&$  0.338$\\
\hline
$3.00^{\rm a}$ &$ -0.12 $&$  0.046 $&$  0.0025 $&$ -0.338$&$  0.000$&$  0.338$\\ \hline
2.50 &$ -0.08 $&$  0.160 $&$  0.0008 $&$  0.000$&$  0.000$&$  0.000$\\
$2.50^{\rm a}$ &$ -0.08 $&$  0.134 $&$  0.0008 $&$ -0.335$&$  0.000$&$  0.335$\\
\hline
$2.50^{\rm a}$ &$ -0.12 $&$  0.114 $&$  0.0020 $&$ -0.338$&$  0.000$&$  0.338$\\
2.50 &$ -0.12 $&$  0.150 $&$  0.0020 $&$  0.000$&$  0.000$&$  0.000$\\
\hline
 $2.00^{\rm a}$ &$ -0.12 $&$  0.180 $&$  0.0020 $&$ -0.342$&$  0.001$&$  0.341$\\
\hline
 $2.00^{\rm a}$ &$ -0.08 $&$  0.200 $&$  0.0008 $&$ -0.338$&$  0.000$&$  0.338$\\
\hline
$1.50^{\rm a}$ &$ -0.08 $&$  0.268 $&$  0.0004 $&$ -0.334$&$  0.000$&$  0.334$\\
\hline
 $1.50^{\rm a}$ &$ -0.12 $&$  0.247 $&$  0.0020 $&$ -0.336$&$  0.001$&$  0.335$\\
\hline
\hline
\end{tabular}

\vspace{0.15 in} \noindent
a) For this line of parameters, adding $-0.002$ to $Q$ takes ${\bf c}$ from
the $(c,-c,0)$ phase into the $(2c,-c,-c)$ phase ({\it e. g.} see the first
and second lines of this table).
\end{table}

\subsection{EXPERIMENTAL DETERMINATION OF PARAMETERS}

In principle we can fix the magnitudes of the dominant exchange
integrals by relating them to several experimentally observed quantities.
These quantities include the value of the ordering temperature, $T_c$, the
Curie-Weiss temperature, $\Theta_{\rm C-W}$, for the susceptibility
\begin{eqnarray}
\chi \sim C/(T-\Theta_{\rm C-W}) \ ,
\label{CW} \end{eqnarray}
 and the high-temperature (compared to $T_c$) specific
heat.\cite{SPHT1,SPHT2} We consider these in turn
and will obtain an estimate for the largest exchange constants $J$
and  $K$ (which here we denote $K_0$ to avoid confusion with
the symbol for Kelvin temperature units.) Crudely we estimate that
due to fluctuations not included within mean field theory
the actual ordering temperature, $3.8$K,
is about $2T_{\rm MF}/3$, so that $T_{\rm MF}\approx 6$K.
From Eq. (\ref{CHIVEQ}) we deduce that (neglecting $L$ and $Q$)
\begin{eqnarray}
24{\rm K} &=& 4T_{\rm MF} = - \lambda ({\bf q})/k \nonumber \\
&=&  (-1.88 K_0 + 0.46J + 5.10M)/k \ ,
\label{TCEQ} \end{eqnarray}
where we took $K_0$ to be negative (bearing in mind the discussion
of Fig. \ref{K}), 
and we evaluated the constants for ${\bf c}=(0.338,-0.338,0)$.
If we only take into account the interaction $K_0$, we
get $K_0/k=-13$ K.  To see what zero-temperature
splitting, $\Delta E$, of the doublet this implies, note that both
$T_{\rm MF}$ and $\Delta E$ are proportional to $J({\bf q})$, the
Fourier transform of the exchange integral.  This type of relation leads to
\begin{eqnarray}
\Delta E = {3 \over S+1} kT_{\rm MF} = 2 kT_{\rm MF}\ ,
\end{eqnarray}
so that $\Delta E /k= 12$K.  This nearly agrees with the result
$\Delta E/k\approx 15$K, given by Boucherle and Schweizer.\cite{PHYSICA}

Next we consider the Curie-Weiss temperature.
This is a particularly good quantity to compare to calculations because,
being the first nontrivial term in the high temperature expansion of the
uniform susceptibility, it is not subject to fluctuation corrections.  
In Appendix B we give a generalization of Eq. (\ref{CW}) which takes
the crystal field splitting into account.  There we show that the
Curie-Weiss intercept extrapolated from values of the susceptibility
$\chi$ at infinite temperature is related to the exchange constants via
\begin{eqnarray}
- \sum_j J_{ij}/k &=& (20/21) \Theta_{\rm C-W} \ .
\label{CWEQ} \end{eqnarray}
Following reference \onlinecite{CURIE1} we set the Curie-Weiss intercept
equal to -33K.  But, as shown in Appendix B, to get this value when an
is made from data at $T<300$K (rather than from infinite temperature),
it is necessary to take
\begin{eqnarray}
- 28.5 {\rm K} &=& \sum_j J_{ij}/k \approx (4K_0 + 12 J)/k \ .
\label{CURIEEQ} \end{eqnarray}
If we neglect $M$, then Eqs. (\ref{TCEQ}) and (\ref{CURIEEQ})
lead to the determination
\begin{eqnarray}
K_0/k &=& -11.3 {\rm K} \ , \ \ \ 
J/k = 6.1 {\rm K} \ .
\end{eqnarray}
The value of $K_0$ is fixed to within about 10\% by Eq.
(\ref{TCEQ}), but the value of $J$ is subject to larger 
(say 20\%) uncertainty.
A question which we can not settle is whether it is justified
to rely on a pure Heisenberg model to interpret that
Curie-Weiss susceptibility.  Attributing contributions to the
susceptibility to the conduction electrons or to the
diamagnetism of core electrons would somewhat
modify our estimates.

The magnetic specific heat $C$ for a system governed by the spin
Hamiltonian ${\cal H}$ gives rise to the limiting value $CT^2/k$ at
infinite temperature given by
\begin{eqnarray}
CT^2 / k &=& {{\rm Tr} {\cal H}^2 \over {\rm Tr} 1}
= {1 \over 2} \sum_{i,j} {{\rm Tr} J_{ij}^2 ({\bf S}_i \cdot {\bf S}_j)^2
\over {\rm Tr} 1 } \nonumber \\ &=&
{1 \over 6} \sum_{i,j} J_{ij}^2 [S(S+1)]^2 =
{3N \over 32} \sum_j J_{ij}^2 \nonumber \\ &=&
{3N \over 32} [4K_0^2 + 12J^2 + 6M^2 + 12 L^2] \ ,
\end{eqnarray}
where $N$ is the total number of Ce ions.  This quantity might
not be easy to determine experimentally because it requires
separating off from the total measured specific heat (in the temperature
range, say, $10{\rm K} < T < 20{\rm K}$), the amount
attributed to the lattice and conduction electrons.

Finally, we should mention that the interactions we determine are
those renormalized by virtual excitation to excited crystal field
states.  Normally, one might ignore such effects.   However,
as we show in Appendix C, the contribution to $J$ from these
virtual processes is of the same order as we have just
determined by our fit to experiment.  These virtual process also
imply that long range interactions must be present even if one
does not invoke RKKY interactions.  So our appeal to the
$Q$ interaction [at separation $(a,a,a)$] is not unreasonable.

\subsection{APPLICATION TO TmS}

At this point we recall that wavevector selection in TmS is of the
same form as for CEAL [see Eq. (\ref{qeq})], but with 
$\delta=0.075$.\cite{TMS2} In TmS the Tm spins form an fcc lattice, so
the lattice geometry is not the same as for CEAL and for TmS the interactions
$K$ and $L$ do not occur.  However, TmS is similar to CEAL in that
one can imagine the dominant exchange interactions limiting one to
be close to the subspace $c_x+c_y+c_z=0$, in which case a major concern
is to have the anisotropy in wavevector space, as in Eq. (\ref{ANISOTROPY}),
so that the incommensuration is of the form ${\bf c} =(\delta, -\delta, 0)$
rather than ${\bf c}=(\delta, \delta, -2 \delta)$. We illustrate this
analogy by a brief numerical survey of the selected wavevector as a function of
the interaction $Q$ (for separation $(1,1,1)$) as in Eq. (\ref{QEQ}).
The result in Table \ref{TMST} shows again the effect of this term on
the anisotropy in wavevector space which can be invoked to explain the
pattern of incommensuration similar to that of CEAL.  In addition, we mention that
like CEAL, no higher harmonics, especially at wavevector $3 {\bf q}$ were
detected.\cite{TMS2}  We propose that, as we show in the next section, this
could be understood if the magnetic structure of TmS were to consist of the
superposition of exactly two wavevectors, as is the case for CEAL.\cite{CEAL5,CEAL6}

\begin{table}
\caption{\label{TMST}Wavevectors at which $\lambda({\bf q})$ is minimal for values
of the exchange interactions (in arbitrary units) for the listed separations (in
units of the lattice constant) on an fcc lattice as a function of the $(1,1,1)$
interaction analogous to $Q$ of Eq. (\protect{\ref{QEQ}}). All interactions are
positive (antiferromagnetic).}

\vspace{0.2 in}
\begin{tabular} {||c c c c c | c c c||} \hline
\multicolumn{5} {|c|} {Exchange Interactions} & \multicolumn{3}
{|c|} {$\cos (q_\alpha a/2)$} \\
\hline
(${1 \over 2},{1 \over 2},0$) & (1,0,0)& (${1 \over 2},{1 \over 2},1$)&
(1,1,0)& (1,1,1)& $c_x$ & $c_y$ & $c_z$ \\
\hline
2.000 & 1.000 & 0.150 & 0.044 &  0.003 & -0.304& 0.000& 0.304\\ 
2.000 & 1.000 & 0.150 & 0.044 &  0.002 & -0.334&-0.002& 0.336\\ 
2.000 & 1.000 & 0.150 & 0.044 &  0.001 & -0.412& 0.150& 0.264\\ 
2.000 & 1.000 & 0.150 & 0.044 &  0.000 & -0.444& 0.224& 0.224\\ 
\hline \end{tabular} \end{table}

\section{QUARTIC TERMS IN THE LANDAU FREE ENERGY}

In this section we analyze the quartic terms in the Landau free energy
in order to investigate the coupling between wavevectors in the star of
${\bf q}$.  Before starting this complicated calculation, we describe
briefly the physical effects we will address.  As the temperature is
lowered in the ordered phase, the effect of the quartic terms in the
Landau free energy, which is to favor fixed length spins, progressively
increases.  This phenomenon is particularly significant for
incommensurate systems.  For many systems having uniaxial anisotropy,
order first occurs in which the spins are aligned along the easy
axis with sinusoidally modulated amplitude.\cite{Nagamiya} In that
case, when the temperature is sufficiently lowered so that the
fourth order terms become important, the fixed length constraint
causes the appearance of transverse spin order, which implies a
phase transition,\cite{Nagamiya} and Ni$_3$V$_2$O$_8$ is a recent example
of this phenomenon.\cite{NVO}  As we shall see, in CEAL the fixed length
constraint favors the simultaneous appearance of incommensurate structures
at the two wavevectors which combine properly to minimize fluctuations
in the spin lengths.  To show this analytically is algebraically
quite complicated, as will become apparent. (If we only wished to show
that the double q state was favored relative to the single q state,
as was done in Ref. \onlinecite{dq2}, the calculation would be much
simpler.  However, our aim was to show that the double q state was
favored over all other possibilities.)

In the preceding section we discussed wavevector selection within a
model of isotropic exchange interactions.  This model is somewhat
misleading in that it has much higher symmetry than that 
required by crystal symmetry.  When more general interactions are present,
the eigenvector of the quadratic free energy matrix associated with
the eigenvalue which first becomes nonpositive as the temperature
is lowered determines the form and symmetry of the long range order.
This critical eigenvector must transform according to an irreducible
representation (irrep) of the symmetry group of the crystal, as is
discussed recently by one of us.\cite{JS1}  This discussion tacitly
assumes the impossibility of accidental degeneracy wherein two or
more irreps having different symmetry could simultaneously condense.
Accordingly we expect that
\begin{eqnarray}
{\bf S}^\tau_\alpha ({\bf R}) &=& \sum_{n=1}^{12} S_\alpha^\tau ({\bf q}_n)
e^{i {\bf q}_n \cdot {\bf R}} \ + {\rm c. \ c.} \ ,
\end{eqnarray}
where ${\bf S}^\tau({\bf R})$ is the spin vector at the $\tau$th
site in the unit cell at ${\bf R}$,
c. c. indicates the complex conjugate of the preceding terms, and
the sum is over the 12 wavevectors ${\bf q}_i$ which, together with
$-{\bf q}_i$, comprise the star of ${\bf q}$.  For some purposes it is
convenient to divide the ${\bf q}_i$'s into three classes
${\bf Q}^\mu_\alpha$ for $\mu=\alpha, \beta, \gamma$ such that for $n=1,2,3,4$
\begin{eqnarray}
{\bf q}_n &=& {\bf Q}^\alpha_n \ , \
{\bf q}_{n+4} = {\bf Q}^\beta_n \ , \
{\bf q}_{n+8} = {\bf Q}^\gamma_n \ ,
\end{eqnarray}
where the ${\bf Q}$'s are listed in Tables \ref{WAVE1}, \ref{WAVE2},
and \ref{WAVE3}. Near the ordering temperature $S_\alpha^\tau({\bf q}_i)$
can be written as a temperature-dependent complex-valued amplitude $x_i(T)$
times the critical eigenvector $m_\alpha^\tau({\bf q}_i)$ normalized by
$\sum_{\alpha n} |m_\alpha^{\tauv_n}|^2 =1$.  Then
\begin{eqnarray}
{\bf S}^\tau_\alpha ({\bf R}) &=& \sum_{n=1}^{12} x_n {\bf m}^\tau ({\bf q}_n)
e^{i{\bf q}_n \cdot {\bf R}} \ + {\rm c. \ c.}
\end{eqnarray}
Thus the $x_n$'s are the complex-valued order parameters of this system.
The result of representation theory for CEAL,
given in Ref. \onlinecite{JS2}, is that for the wavevector
${\bf q}_1=(1/2-\delta,1/2+\delta,1/2)(2 \pi /a)$,
the critical eigenvector, which gives the spin components of the two
sites in the unit cell for the
irrep which experiments\cite{CEAL1,CEAL2,CEAL5,CEAL6}  have shown to
be the active one, is of the form
\begin{eqnarray}
{\bf m}^{\tauv_1}({\bf q}_1)&=& (\alpha e^{i \phi} , \alpha e^{-i\phi} \ ,
\beta) \ , \nonumber \\
{\bf m}^{\tauv_2}({\bf q}_1) &=& (-\alpha e^{-i \phi} , -\alpha e^{i\phi} \ ,
\beta) \ ,
\label{EQ99} \end{eqnarray}
where the real-valued parameters $\alpha$, $\beta$, and $\phi$ depend on
the interactions but can be determined from experimental data.
The next step in this calculation is to use crystal symmetry to relate
the eigenvectors for the other wavevectors in the star of ${\bf q}$
to that given in Eq. (\ref{EQ99}).  This
is done in the Appendix and the results are listed in Tables
\ref{WAVE1}, \ref{WAVE2}, and \ref{WAVE3}.  

\vspace{0.2 in}
\begin{table*} [h]
\caption{\label{WAVE1} Wavefunctions for the ${\bf Q}^\alpha$
wavevectors.$^{(a)}$}
\vspace{0.2 in}
\begin{tabular} {||c | c | c | c || c | c | c || c | c | c ||}
\hline \hline
$n$ & \multicolumn {3} {|c||} {${\bf Q}^\alpha_n$} &
\multicolumn{3} {|c||} {${\bf m}^1({\bf Q}^\alpha_n)$} &
\multicolumn{3} {|c||} {${\bf m}^2({\bf Q}^\alpha_n)$} \\
\hline
1 & ${1 \over 2} - \delta$ & ${1 \over 2} + \delta$ &  ${1 \over 2} $
& $\alpha e^{i\phi}$ & $\alpha e^{-i\phi}$ & $\beta$ & $-\alpha e^{-i\phi}$
& $-\alpha e^{i\phi}$ & $-\beta$ \\
2 & ${1 \over 2} - \delta$ & ${1 \over 2} + \delta$ &  $-{1 \over 2}$
& $\alpha e^{-i\phi}$ & $\alpha e^{i\phi}$ & $-\beta$ & $\alpha e^{i\phi}$
& $\alpha e^{-i\phi}$ & $-\beta$ \\
3 & ${1 \over 2} - \delta$ & $-{1 \over 2} - \delta$ & ${1 \over 2}$ &
$-\alpha e^{i(\pi \delta - \phi)}$ & $\alpha e^{i(\pi \delta + \phi)}$ &
$-\beta e^{i\pi \delta}$ & $-\alpha e^{i\phi}$ & $\alpha e^{-i\phi}$ &
$-\beta$ \\
4 & ${1 \over 2} - \delta$ & $-{1 \over 2} - \delta$ & $-{1 \over 2}$
& $-\alpha e^{i(\pi \delta+ \phi) } $ & $\alpha e^{i(\pi \delta- \phi) }$ 
& $\beta e^{i \pi \delta}$ & $-\alpha e^{-i\phi}$ & $\alpha e^{i\phi}$ &
$\beta$ \\ 
\hline \hline
\end{tabular}

\vspace{0.2 in}
\noindent a) Wavevectors are given in units of $2\pi /a$.
\end{table*}

\vspace{0.2 in}
\begin{table*} [h]
\caption{\label{WAVE2} Wavefunctions for the ${\bf Q}^\beta$
wavevectors.$^{(a)}$}
\vspace{0.2 in}
\begin{tabular} {||c | c | c | c || c | c | c || c | c | c ||}
\hline \hline
$n$ & \multicolumn {3} {|c||} {${\bf Q}^\beta_n$} &
\multicolumn{3} {|c||} {${\bf m}^1({\bf Q}^\beta_n)$} &
\multicolumn{3} {|c||} {${\bf m}^2({\bf Q}^\beta_n)$} \\
\hline
1 & ${1 \over 2}$ & ${1 \over 2} - \delta$ & ${1 \over 2} + \delta$
& $\beta$ & $\alpha e^{i\phi}$ & $\alpha e^{-i\phi}$ & $-\beta$ &
$-\alpha e^{-i\phi}$ & $-\alpha e^{i\phi}$ \\
2 & $-{1 \over 2}$ & ${1 \over 2} - \delta$ & ${1 \over 2} + \delta$
&$-\beta$ &  $\alpha e^{-i\phi}$ & $\alpha e^{i\phi}$ & $-\beta$ &
$\alpha e^{i\phi}$ & $\alpha e^{-i\phi}$ \\
3 & ${1 \over 2}$ & ${1 \over 2} - \delta$ & $-{1 \over 2} - \delta$
& $-\beta e^{i \pi \delta}$ & $-\alpha e^{i(\pi \delta-\phi)}$ &
$\alpha e^{i(\pi \delta+\phi)}$ & $-\beta$ & $-\alpha e^{i\phi}$
& $\alpha e^{-i\phi}$ \\
4 & $-{1 \over 2}$ & ${1 \over 2} - \delta$ & $-{1 \over 2} - \delta$
& $\beta e^{i\pi \delta}$ & $-\alpha e^{i(\pi \delta +\phi)}$ &
$\alpha e^{i(\pi \delta - \phi)}$
& $\beta$ & $-\alpha e^{-i\phi}$ & $\alpha e^{i\phi}$ \\ 
\hline \hline
\end{tabular}

\vspace{0.2 in}
\noindent a) Wavevectors are given in units of $2\pi /a$.
\end{table*}

\vspace{0.2 in}
\begin{table*} [h]
\caption{\label{WAVE3} Wavefunctions for the ${\bf Q}^\gamma$
wavevectors.$^{(a)}$}
\vspace{0.2 in}
\begin{tabular} {||c | c | c | c || c | c | c || c | c | c ||}
\hline \hline
$n$ & \multicolumn {3} {|c||} {${\bf Q}^\gamma_n$} &
\multicolumn{3} {|c||} {${\bf m}^1({\bf Q}^\gamma_n)$} &
\multicolumn{3} {|c||} {${\bf m}^2({\bf Q}^\gamma_n)$} \\
\hline
1 & ${1 \over 2} + \delta$ & $ {1 \over 2}$ & ${1 \over 2} - \delta$
& $\alpha e^{-i\phi}$ & $\beta$ & $\alpha e^{i\phi}$ & $-\alpha e^{i\phi}$
& $-\beta$ & $-\alpha e^{-i\phi}$ \\
2 & ${1 \over 2} + \delta$ & $-{1 \over 2}$ & ${1 \over 2} - \delta$
& $\alpha e^{i\phi}$ & $-\beta$ & $\alpha e^{-i\phi}$ & $\alpha e^{-i\phi}$
& $-\beta$ & $\alpha e^{i\phi}$ \\
3 & $-{1 \over 2} - \delta$ & ${1 \over 2}$ & ${1 \over 2}-\delta$
& $\alpha e^{i(\pi \delta+\phi)}$ & $-\beta e^{i\pi \delta}$ &
$-\alpha e^{i(\pi \delta-\phi)}$ & $\alpha e^{-i\phi}$ & 
$-\beta$ & $-\alpha e^{i\phi}$ \\
4 &$-{1 \over 2} - \delta$ & $-{1 \over 2}$ & ${1 \over 2} - \delta$
& $\alpha e^{i(\pi \delta-\phi)}$ & $\beta e^{i \pi \delta}$
& $-\alpha e^{i(\pi \delta +\phi)}$ & $\alpha e^{i\phi}$ & $\beta$ &
$-\alpha e^{-i\phi}$ \\ 
\hline \hline
\end{tabular}

\vspace{0.2 in}
\noindent a) Wavevectors are given in units of $2\pi /a$.
\end{table*}

We now turn to the calculation. Equation (\ref{TSEQ}) shows that
the fourth order terms in the Landau free energy are
\begin{eqnarray}
F_4 &=& N_{\rm uc}^{-1} bkT \sum_{{\bf R}, \tau}
[{\bf S}^\tau ({\bf R}) \cdot {\bf S}^\tau ({\bf R}) ]^2 \ ,
\end{eqnarray}
where $b$ is a constant of order unity (henceforth we set $bkT=1$).
In terms of the order parameters $x_i$ the free energy per unit cell is
\begin{eqnarray}
F &=& \chi^{-1} \sum_{i=1}^{12} |x_i|^2 + F_4 \ ,
\end{eqnarray}
where $\chi^{-1} = 4kT + \lambda({\bf q}_1)$ when the small perturbations
to the isotropic Heisenberg model are ignored.  At quadratic order,
there is complete isotropy within the order parameter space of twelve
complex variables.  Our objective is to find the direction in the
space of the $x$'s which has the lowest free energy.  This direction
will indicate whether condensation (when ordering takes place) takes
place via a single wavevector or via the simultaneous condensation into
more than
one wavevector.  To study this anisotropy, we will consider the subspace
\begin{eqnarray}
\sum_n |x_n|^2 &=& c \ ,
\end{eqnarray}
where we take $c=1$, for convenience.  We write
\begin{eqnarray}
F_4  &=& N_{uc}^{-1} \sum_{{\bf R} \tau}
\sum_{q_1,q_2,q_3,q_4} \sum_{\alpha \beta}
S^\tau_\alpha(q_1) S^\tau_\alpha(q_2) S^\tau_\beta(q_3)
S^\tau_\beta(q_4) \nonumber \\ && \ \times
\exp [i({\bf q_1} + {\bf q_2} + {\bf q_3}
+ {\bf q_4) \cdot {\bf R}}] \nonumber \\ &=& \sum_{{\bf G} \tau}
\sum_{q_1,q_2,q_3,q_4} \sum_{\alpha \beta} S^\tau_\alpha(q_1)
S^\tau_\alpha(q_2) S^\tau_\beta(q_3) S^\tau_\beta(q_4)
\nonumber \\ && \ \times
\delta_{{\bf G},{\bf q_1} + {\bf q_2} + {\bf q_3} + {\bf q_4}} \ ,
\end{eqnarray}
where the delta function conserves wavevector to within a reciprocal
lattice vector ${\bf G}$.  

We will decompose $F_4$ into terms involving different sets of the
critical wavevectors  ${\bf q}_n$  (and their negatives) and will
express the results in terms of the order parameters $x_n$.  We write
\begin{eqnarray}
F_4 &=& \sum {\cal S} \ .
\end{eqnarray}
The first set of terms which we consider are those which involve only one
wavevector ${\bf q}$ (By this kind of statement we always mean ${\bf q}$ and
$-{\bf q}$.) which we denote ${\cal S}_1$, where
\begin{eqnarray}
{\cal S}_1 &=& \sum_{i=1}^{12} |x_i|^4 \sum_\tau \left[
2 \biggl| \sum_\alpha m_\alpha^\tau ({\bf q}_i)^2 \biggr|^2
\right. \nonumber \\ && \ + \left. 4 \left( \sum_\alpha \biggl|
m_\alpha^\tau({\bf q}_i) \biggr|^2 \right)^2 \right] \ .
\end{eqnarray}

Next we consider terms involving exactly two different wavevectors
${\bf q}_i$ and ${\bf q}_j$.  These are of two kinds, which we denote
${\cal S}_{2a}$ and ${\cal S}_{2b}$.  In the first of these
we automatically conserve wavevector by taking pairs of opposite
wavevectors.  This term (which occurs for arbitrarily chosen pairs of
wavevectors) is
\begin{eqnarray}
{\cal S}_{2a} &=& 8 \sum_{i < j} |x_i|^2 |x_j|^2 \sum_\tau \Biggl\{ 
\left( \sum_\alpha \biggl| m_\alpha^\tau ( {\bf q}_i) \biggr|^2 \right)
\nonumber \\ && \ \times
\left( \sum_\beta \biggl| m_\beta^\tau ( {\bf q}_j) \biggr|^2 \right)
+  \Biggl| \sum_\alpha m_\alpha^\tau ( {\bf q}_i) m_\alpha^\tau({\bf q}_j)
\Biggr|^2 \nonumber \\ && \ +  \Biggl| \sum_\alpha 
m_\alpha^\tau ( {\bf q}_i) m_\alpha^\tau(-{\bf q}_j) \Biggr|^2 \Biggr\} \ .
\end{eqnarray}
The second kind of term is one in which $2q_i-2q_j$ is equal to
a nonzero reciprocal lattice vector, ${\bf G}$.  This term is
\begin{eqnarray}
{\cal S}_{2b} &=& \sum_{i\not= j} x_i^2 {x_j^*}^2 \sum_\tau
\sum_{{\bf G}\not= 0} \delta_{2{\bf q}_i - 2{\bf q}_j, {\bf G}}
\nonumber \\ &\times & \Biggl\{ 2 \left( \sum_\alpha
m^\tau_\alpha({\bf q}_i)^2 \right) \left(  \sum_\beta 
m^\tau_\beta(-{\bf q}_j)^2 \right) \nonumber \\ &+& 4 \left(
\sum_\alpha m^\tau_\alpha({\bf q}_i) m^\tau_\alpha(-{\bf q}_j) \right)^2
\Biggr\} \ .
\end{eqnarray}
Wavevector conservation in these terms is only satisfied when the
two wavevectors involved are $q_{2n-1}$ and $q_{2n}$ and it is
exactly this pair of wavefunctions that are coupled in the
observed ``double-q'' state.\cite{CEAL5,CEAL6}

There are no terms involving exactly three distinct wavevectors.
The terms involving four wavevectors, denoted ${\cal S}_{4,m}$,
involve the wavevectors
\begin{eqnarray}
{\cal S}_{4,1}: && ({\bf q}_1, -{\bf q}_3, {\bf q}_5, {\bf q}_8) \nonumber \\
{\cal S}_{4,2}: && ({\bf q}_1, -{\bf q}_3, {\bf q}_6, {\bf q}_7) \nonumber \\
{\cal S}_{4,3}: && ({\bf q}_2, -{\bf q}_4, {\bf q}_5, {\bf q}_8) \nonumber \\
{\cal S}_{4,4}: && ({\bf q}_2, -{\bf q}_4, {\bf q}_6, {\bf q}_7) \nonumber \\
{\cal S}_{4,5}: && ({\bf q}_1, -{\bf q}_2, {\bf q}_7, -{\bf q}_8) \nonumber \\
{\cal S}_{4,6}: && ({\bf q}_1, -{\bf q}_2, -{\bf q}_7, {\bf q}_8) \ ,
\label{TEQ} \end{eqnarray}
the negatives of these, and the set of wavevectors obtained by
the permutation ${\bf Q}^\alpha \rightarrow
{\bf Q}^\beta \rightarrow {\bf Q}^\gamma \rightarrow {\bf Q}^\alpha$,
which amounts to $q_n \rightarrow q_{n+4}$. Here and below
$n+4$ is interpreted as $n-8$ when $n+4$ is greater than 12.
We used a computer program to check that the terms we have enumerated
are the only ones which can appear in fourth order.
We write out the first of these:
\begin{eqnarray}
{\cal S}_{4,1} &=& 8 x_1x_3^*x_5x_8 \nonumber \\
&\times & \sum_\tau \Biggl[ \biggl( {\bf m}^\tau ( {\bf Q}_1^\alpha ) \cdot
{\bf m}^\tau ( {\bf Q}_3^\alpha )^* \biggr) \biggl(
{\bf m}^\tau ( {\bf Q}_1^\beta ) \cdot {\bf m}^\tau ( {\bf Q}_4^\beta )
\biggr) 
\nonumber \\ &+& \biggl( {\bf m}^\tau ( {\bf Q}_1^\alpha ) \cdot
{\bf m}^\tau ( {\bf Q}_1^\beta ) \biggr) \biggl(
{\bf m}^\tau ( {\bf Q}_3^\alpha )^* \cdot {\bf m}^\tau ( {\bf Q}_4^\beta )
\biggr) \nonumber \\ &+& \biggl( {\bf m}^\tau ( {\bf Q}_1^\alpha ) \cdot
{\bf m}^\tau ( {\bf Q}_4^\beta ) \biggr) \biggl(
{\bf m}^\tau ( {\bf Q}_3^\alpha )^* \cdot {\bf m}^\tau ( {\bf Q}_1^\beta )
\Biggr]
\end{eqnarray}

\begin{table} [h]
\caption{\label{wave1} Wavefunctions for each wavevector.
We list $\sqrt 6 m^\tau_\alpha$.}
\vspace{0.2 in}
\begin{tabular} {||c | c | c | c || c | c | c || c | c | c ||}
\hline \hline
$n$ & \multicolumn {3} {|c||} {${\bf q}_n$} &
\multicolumn{3} {|c||} {${\bf m}^1({\bf q}_n)$} &
\multicolumn{3} {|c||} {${\bf m}^2({\bf q}_n)$} \\
\hline
1 & ${1 \over 2} - \delta$ & ${1 \over 2} + \delta$ &  ${1 \over 2} $
& $1$ & $1$ & $1$ & $-1$ & $-1$ & $-1$ \\
2 & ${1 \over 2} - \delta$ & ${1 \over 2} + \delta$ &  $-{1 \over 2}$
& $1$ & $1$ & $-1$ & $1$ & $1$ & $-1$ \\
3 & ${1 \over 2} - \delta$ & $-{1 \over 2} - \delta$ & ${1 \over 2}$ &
$-e^{i\pi \delta}$ & $e^{i\pi \delta}$ & 
$-e^{i\pi \delta}$ & $-1$ & $1$ & $-1$ \\
4 & ${1 \over 2} - \delta$ & $-{1 \over 2} - \delta$ & $-{1 \over 2}$ &
$-e^{i\pi \delta}$ & $e^{i\pi \delta}$ & 
$e^{i\pi \delta}$ & $-1$ & $1$ & $1$ \\
\hline
5 & ${1 \over 2}$ & ${1 \over 2} - \delta$ & ${1 \over 2} + \delta$
& $1$ & $1$ & $1$ & $-1$ & $-1$ & $-1$ \\
6 & $-{1 \over 2}$ & ${1 \over 2} - \delta$ & ${1 \over 2} + \delta$
& $-1$ & $1$ & $1$ & $-1$ & $1$ & $1$ \\
7 & ${1 \over 2}$ & ${1 \over 2} - \delta$ & $-{1 \over 2} - \delta$ &
$-e^{i\pi \delta}$ & $-e^{i\pi \delta}$ & 
$e^{i\pi \delta}$ & $-1$ & $-1$ & $1$ \\
8 & $-{1 \over 2}$ & ${1 \over 2} - \delta$ & $-{1 \over 2} - \delta$ &
$e^{i\pi \delta}$ & $-e^{i\pi \delta}$ & 
$e^{i\pi \delta}$ & $1$ & $-1$ & $1$ \\
\hline
9 & ${1 \over 2} + \delta$ & $ {1 \over 2}$ & ${1 \over 2} - \delta$
& $1$ & $1$ & $1$ & $-1$ & $-1$ & $-1$ \\
10 & ${1 \over 2} + \delta$ & $-{1 \over 2}$ & ${1 \over 2} - \delta$
& $1$ & $-1$ & $1$ & $1$ & $-1$ & $1$ \\
11 & $-{1 \over 2} - \delta$ & ${1 \over 2}$ & ${1 \over 2}-\delta$ &
$e^{i\pi \delta}$ & $-e^{i\pi \delta}$ & 
$-e^{i\pi \delta}$ & $1$ & $-1$ & $-1$ \\
12 &$-{1 \over 2} - \delta$ & $-{1 \over 2}$ & ${1 \over 2} - \delta$ &
$e^{i\pi \delta}$ & $e^{i\pi \delta}$ & 
$-e^{i\pi \delta}$ & $1$ & $1$ & $-1$ \\
\hline \hline
\end{tabular}
\end{table}

We will now treat the case applicable to CEAL when ${\bf m}^\tau ({\bf q}_n)$
is parallel to the appropriate (1,1,1) direction,\cite{CEAL1,CEAL2} in which
case the wavefunctions are those given in Table
\ref{wave1}.  Note that whenever a ${\bf m}({\bf Q}^{\, \rho}_3)$ or
${\bf m}({\bf Q}^{\, \rho}_4)$ appears in one of these fourth order
terms, then a ${\bf m}({\bf Q}^{\, \rho}_4)^*$ or 
${\bf m}({\bf Q}^{\, \rho}_3)^*$ also appears.  This means that in using
the wavefunctions, we may replace $e^{i \pi \delta}$ by unity.
Also note that the wavefunction for ${\bf Q}^{\, \rho}_n$ changes sign for
$n=1$ on going from $\tau=1$ to $\tau=2$, whereas the wavefunctions for
$n \not= 1$ do not change sign. This means that any term which contains an
odd number of variables $Q^\alpha_n$ with $n=1$ vanishes when the sum over
$\tau$ is performed.  Thus, out of those terms listed above,
only ${\cal S}_{4,1}$ and ${\cal S}_{4,4}$ (their negative and their
cyclically permuted partners) survive the sum over $\tau$. 
We will also need  (for $\tau=1$ and $\delta=0$)
\begin{eqnarray}
M_{ij} &\equiv& 6 \sum_\alpha m_\alpha^\tau ({\bf q}_i)
m_\alpha^\tau ({\bf q}_j) \nonumber \\
& =& 6 \sum_\alpha m_\alpha^\tau ({\bf q}_i)
m_\alpha^\tau (-{\bf q}_j) \ ,
\label{MIJEQ} \end{eqnarray}
which we list in Table \ref{MATRIX}.

\begin{table}
\caption{\label{MATRIX} Matrix elements $M_{ij}$ of Eq. 
(\protect{\ref{MIJEQ}}).}
\begin{tabular} {||c | c | c | c | c | c | c | c | c | c | c | c | c ||}
\hline
$i/j$ & 1 & 2 & 3 & 4 & 5 & 6 & 7 & 8 & 9 & 10 & 11 & 12\\
\hline
1& 3 & 1 & $-1$ & 1 & 3 & 1 & $-1$ & 1 & 3 & 1 & $-1$ & 1 \\
2& 1 & 3 & $1$ & $-1$ & 1 & $-1$ & $-3$ & $-1$ & 1 & $-1$ & $1$ & 3 \\
3& $-1$ & 1 & $3$ & $1$ & $-1$ & $1$ & $-1$ & $-3$ & $-1$ & $-3$ & $-1$ & 1 \\
4& $1$ & $-1$ & $1$ & $3$ & $1$ & $3$ & $1$ & $-1$ & $1$ & $-1$ & $-3$ & $-1$ \\
\hline
5& $3$ & $1$ & $-1$ & $1$ & $3$ & $1$ & $-1$ & $1$ & $3$ & $1$ & $-1$ & $1$ \\
6&$1$ & $-1$ & $1$ & $3$ & $1$ & $3$ & $1$ & $-1$ & $1$ & $-1$ & $-3$ & $-1$ \\
7&$-1$ & $-3$ & $-1$ & $1$ & $-1$ & $1$ & $3$ & $1$ & $-1$ & $1$ & $-1$ & $3$ \\
8& $1$ & $-1$ & $-3$ & $-1$ & $1$ & $-1$ & $1$ & $3$ & $1$ & $3$ & $1$ & $-1$ \\
\hline
9& $3$ & $1$ & $-1$ & $1$ & $3$ & $1$ & $-1$ & $1$ & $3$ & $1$ & $-1$ & $1$ \\
10&$1$ & $-1$ & $-3$ & $-1$ & $1$ & $-1$ & $1$ & $3$ & $1$ & $3$ & $1$ & $-1$ \\
11&$-1$&$1$ & $-1$ & $-3$ & $-1$ & $-3$ & $-1$ & $1$ & $-1$ & $1$ & $3$ & $1$ \\
12& $1$ & $3$ & $1$ & $-1$ & $1$ & $-1$ & $3$ & $-1$ & $1$ & $-1$ & $1$ & $3$ \\
\hline
\end{tabular}
\end{table}

Thus we have the result
\begin{eqnarray}
&& {\cal S}_1 + {\cal S}_{2a} + {\cal S}_{2b} =
2 + \sum_{i=1}^{12} \left| x_i \right|^4
+ {8 \over 9} \sum_{i<j} \left| x_i^2 x_j^2 \right|^2 M_{ij}^2
\nonumber \\ && \
+ {11 \over 9} \sum_{n=1}^6 \Biggl[ x_{2n-1}^2 [x_{2n}^*]^2
+ x_{2n}^2 [x_{2n-1}^*]^2 \Biggr] \ ,
\label{S12EQ} \end{eqnarray}
and
\begin{eqnarray}
&& \ \sum_n {\cal S}_{4n} = {4 \over 9}
\biggl[ ( M_{13} M_{58} + M_{15} M_{38} + M_{18} M_{53} )
x_1 x_3^* x_5 x_8 \nonumber \\ 
&+& ( M_{24} M_{67} + M_{26} M_{47} + M_{27} M_{46} )
x_2 x_4^* x_6 x_7 \nonumber \\ 
&+& ( M_{57} M_{9,12} + M_{59} M_{7,12} + M_{5,12} M_{79} )
x_5 x_7^* x_9 x_{12} \nonumber \\ 
&+& ( M_{68} M_{10,11} + M_{6,10} M_{8,11} + M_{6,11} M_{8,10} )
x_6 x_8^* x_{10} x_{11} \nonumber \\ 
&+& ( M_{9,11} M_{14} + M_{91} M_{11,4} + M_{94} M_{11,1} )
x_9 x_{11}^* x_1 x_4 \nonumber \\ 
&+& ( M_{10,12} M_{23} + M_{10,2} M_{12,3} + M_{10,3} M_{12,2} )
x_{10} x_{12}^* x_2 x_3 ) \biggr] \nonumber \\ && \ + {\rm c. \ c.}
\nonumber \\ &=& - {44 \over 9} \biggl[ \sum_{n=0}^2
x_{4n+1} x_{4n+3}^* x_{4n+5} x_{4n+8}
\nonumber \\ && \ 
+ x_{4n+2} x_{4n+4}^* x_{4n+6} x_{4n+7} \biggr] + {\rm c. \ c. } \ ,
\end{eqnarray}
where here and below the index $4n+k$ is interpreted as $4n+k-12$ if it
is greater than 12.  We minimize ${\cal S}$ by fixing the phases optimally,
{\it i. e.} so that
\begin{eqnarray}
x_{2n-1} &=& e^{-i \pi /4} r_{2n-1} \ , \ \ \ \ \
x_{2n} = e^{i \pi /4} r_{2n} \ ,
\end{eqnarray}
where all the $r$'s are real and nonnegative.  Then
\begin{eqnarray}
{\cal S} &=& 2 + \sum_{i=1}^{12} r_i^4 
- {22 \over 9} \sum_{n=1}^6 r_{2n-1}^2 r_{2n}^2
+ {8 \over 9} \sum_{i<j} r_i^2 r_j^2 \nonumber \\ && \
+ {64 \over 9} \Biggl[ r_1^2 (r_5^2 + r_9^2)
+ r_2^2(r_7^2+r_{12}^2) + r_3^2(r_8^2+r_{10}^2) \nonumber \\ && \
+ r_4^2(r_6^2+r_{11}^2)
+ r_5^2r_9^2+r_6^2 r_{11}^2+r_7^2r_{12}^2 + r_8^2r_{10}^2 \Biggr] 
\nonumber \\ && \ - {88 \over 9} \Biggl[
r_1 r_3 r_5 r_8 + r_2 r_4 r_6 r_7 + r_5 r_7 r_9 r_{12}
\nonumber \\ && \ + r_6 r_8 r_{10} r_{11} + r_9 r_{11} r_1 r_4
+ r_{10} r_{12} r_2 r_3 \Biggr] \ .
\end{eqnarray}
This is to be minimized under the constraint
\begin{eqnarray}
\sum_{i=1}^{12} r_i^2  &=& 1 \ .
\end{eqnarray}
To do this write ${\cal S} = {\cal S}_A + {\cal S}_B$, where
\begin{eqnarray}
{\cal S}_A &=& 2 + \sum_{k=1}^6 \biggl( r_{2k-1}^2 - r_k^2 \biggr)^2
+ {4 \over 9} \sum_{k=1}^6 r_{2k-1}^2 r_{2k}^2 \nonumber \\ && \
+ {8 \over 9} {\sum_{i < j}}^\prime r_i^2 r_j^2 \ ,
\end{eqnarray}
where the prime on the summation means that we omit terms for which
$i=2k-1$ and $j=2k$, and
\begin{eqnarray}
{\cal S}_B &=& {44 \over 9} \sum_{n=0}^2
\Biggl[ (r_{4n+1}r_{4n+5}-r_{4n+3}r_{4n+8})^2 \nonumber \\ && \
 + (r_{4n+2}r_{4n+7}-r_{4n+4}r_{4n+6})^2 \nonumber \\ && \
+ {20 \over 9}
\sum_{n=0}^2 \biggl[ r_{4n+1}^2 r_{4n+5}^2 + r_{4n+2}^2r_{4n+7}^2 
\nonumber \\ && \ + r_{4n+3}^2r_{4n+8}^2 + r_{4n+4}^2r_{4n+6}^2\Biggr] \ .
\end{eqnarray}
We will minimize ${\cal S}_A$ with respect to the $r_i$'s. 
For the set of $r_i$'s that minimize ${\cal S}_A$, it will happen that
the nonnegative quantity ${\cal S}_B$ is zero.  This shows that this set
of $r_i$'s minimizes ${\cal S}$.

To minimize ${\cal S}_A$ we handle the constraint by introducing a Lagrange
parameter $2 \lambda$.  Then the equations which locate extrema of ${\cal S}_A$,
namely $\partial {\cal S}_A / \partial r_n - 4 \lambda r_n=0$,
are (for $n=1,2,3,4,5,6$)
\begin{eqnarray}
4 r_{2n-1} \Biggl[ {5 \over 9} r_{2n-1}^2 - {11 \over 9} r_{2n}^2
+ {4 \over 9} \sum_{k=1}^{12}  r_k^2 - \lambda \Biggr] = 0 \ , \nonumber \\
4 r_{2n} \Biggl[ - {11 \over 9} r_{2n-1}^2 + {5 \over 9} r_{2n}^2
+ {4 \over 9} \sum_{k=1}^{12}  r_k^2 - \lambda \Biggr] = 0 \ .
\end{eqnarray} 
If both $r_{2n-1}$ and $r_{2n}$ are nonzero, then by subtracting their
equations, one obtains
\begin{eqnarray}
(16/9) (r_{2n-1}^2 - r_{2n}^2 ) = 0 \ .
\end{eqnarray}
Thus $r_{2n-1}^2=r_{2n}^2$ and since $r_n$ is nonnegative, we set
\begin{eqnarray}
r_{2n-1}=r_{2n}=X_n \ .
\end{eqnarray}
Now consider $X_n$ and $X_m$.  Add equations for $r_{2n-1}$ and $r_{2n}$
and subtract those for $r_{2m-1}$ and $r_{2m}$.  Thereby one obtains
\begin{eqnarray}
- {2 \over 3} ( r_{2n-1}^2 + r_{2n}^2)
+ {2 \over 3} ( r_{2m-1}^2 + r_{2m}^2) = 0 \ ,
\end{eqnarray}
which indicates 
that $X_n^2=X_m^2$. So for all pairs $r_{2n-1}$, $r_{2n}$ both of whose
members are nonzero, we may set their $X$'s all equal to $X$, say. In a
similar fashion we show that for all such pairs which have only one nonzero
member we may set the nonzero member equal to $Y$, the same for all such
singly nonzero pairs.  So we characterize the minimum as having
$k$ pairs of doubly nonzero members, each with value $X$, and $l$ pairs
of singly nonzero members assuming the value $Y$.  Then we have that
\begin{eqnarray}
{\cal S}_A &=& 2+ l Y^4 + {4 \over 9} k X^4 + {8 \over 9} \Biggl[
2k(k-1)X^4 \nonumber \\ && \ + 2kl X^2Y^2 + (l/2)(l-1)Y^4 \Biggr] \ ,
\end{eqnarray}
with the constraint
\begin{eqnarray}
2kX^2 + lY^2 = 1 \ .
\end{eqnarray}
This leads to the result that
\begin{eqnarray}
{\cal S}_A &=& 2 + {1 \over l} \biggl( 1 - 2kX^2\biggr)^2 + {4 \over 9} kX^4
+{8 \over 9} \Biggl[ 2k(k-1)X^4 \nonumber \\ && \  + 2kX^2(1-2kX^2)
+ {l-1 \over 2l} (1-2kX^2)^2 \Biggr] \nonumber \\
&\equiv& A X^4 + BX^2 + C \ ,
\end{eqnarray}
where
\begin{eqnarray}
A &=& {4k^2 \over l} + {4 \over 9} k + {16 \over 9} k(k-1) - {32 \over 9}
k^2 + {16 \over 9l} k^2(l-1) \nonumber \\ &=& 20k^2/(9l) -4k/3 \ , 
\end{eqnarray}
\begin{eqnarray}
B &=& -{4k \over l} + {16k \over 9} - {16k(l-1) \over 9l} = - 20 k/(9l) \ ,
\end{eqnarray}
and
\begin{eqnarray}
C &=& 2 + {1 \over l} + {4(l-1) \over 9l} = {5 + 22l \over 9l} \ .
\end{eqnarray}
If 
\begin{eqnarray}
-B/(2A) < X_{\rm max}^2 = 1/(2k) \ ,
\label{LESSEQ} \end{eqnarray}
then the quadratic form is minimized by
setting $X^2=-B/(2A)$. Otherwise, the minimum is realized for
$X=X_{\rm max}$ (for which $l=0$).  We see that we never have the case
of Eq. (\ref{LESSEQ}) because
\begin{eqnarray}
 - {B \over A} = {20k/(9l) \over 20 k^2/(9l) -4k/3} = {1 \over k - \Delta}
\geq {1 \over k} \ .
\end{eqnarray}
Therefore the minimum occurs for
$X=X_{\rm max}$ and $l=0$, where
\begin{eqnarray}
{\cal S}_A &=& {1 \over 36lk^2} \Biggl[ 9lA + 18lkB + 36lk^2C\Biggr]
\nonumber \\ &=&
{1 \over 36lk^2} \Biggl[ 20k^2 -12kl -40k^2 +20k^2 + 88k^2l \Biggr]
\nonumber \\ &=& {22 \over 9} - {1 \over 3k} \ .
\end{eqnarray}
So we conclude that the minima occur for $k=1$, and for
only $r_{2n-1}$ and $r_{2n}$ nonzero, one sees that ${\cal S}_B=0$,
so that the minima of ${\cal S}_A$ are indeed the minima of ${\cal S}$.
These minima correspond to exactly what we want: a single pair of equal
amplitude order parameters of the type we hoped for.

It should also be noted that the phase difference\cite{LOCK} between the two
condensed waves, given by $x_{2n}/x_{2n-1}=e^{i \pi /2}$ also
agrees with the conclusions of Forgan {\it et al.}\cite{CEAL5}
that the structures of the two incommensurate wavevectors add
in quadrature.  In addition, our calculation supports their argument
that the variation of the magnitude of the spin over the incommensurate
wave should be minimial.  Our calculation also explains why the
fixed length constraint does not require substantial values of higher
harmonics, such as ${\bf S}(3{\bf q})$.  However, this picture can not be
totally correct, because the double-q structure does not
completely eliminate the variation of the magnitude of the spin.
The spin structure consists of two helices of opposite chirality and
the ellipticity of these helices decreases with decreasing 
temperature, but the eccentricity of the polarization ellipse
extrapolated to zero temperature\cite{CEAL6} is too large
to be explained by anisotropy alone.  Probably some, or all, of this
eccentricity should be explained by Kondo-like behavior.\cite{CEAL6}

\section{CONCLUSION}

We may summarize our conclusions as follows.

\begin{itemize}

\item
For the fcc antiferromagnet with first and second neighbor interactions we
located a previously overlooked multiphase point [see Eq. (\ref{MULTIEQ})]
at which wavevector selection is infinitely degenerate.

\item
We have extended the analysis of Yamamoto and Nagamiya\cite{YN}
to determine the minimum free energy of magnetic structures
of CeAl$_2$ (which is a two sublattice fcc incommensurate magnet)
for a model consisting of three shells of isotropic exchange
interactions.  The phase diagram in terms of these interactions 
(see Figs. \ref{MINM} and \ref{MINP}) has
an incommensurate phase with a wavevector in a degenerate manifold
which includes the observed incommensurate wavevector for CeAl$_2$.

\item
We analyzed the effect of third nearest neighbors on the
degenerate manifold of the three shell model and found that
it gave the wrong anisotropy in wavevector space to explain
the data for CeAl$_2$.  However, the correct sign of the anisotropy
(which would give a wavevector of the form $(1/2-\delta, 1/2+\delta, 1/2)$
in units of $2 \pi /a$), can be obtained if the interaction $Q$ of
neighbors at separation $(a,a,a)$ exceeds a rather small threshold
value.  Since CeAl$_2$ is a metal subject to RKKY\cite{RK} interactions,
 we suggest that such an interaction is not unreasonable. By way of
illustration we give (see Table \ref{EXTABLE}).
some explicit values of exchange parameters that will give
the correct incommensurate wavevectors.

\item
By analyzing the form of the fourth order terms in the Landau expansion,
we show that for the wavevectors appropriate to the ordered phase of
CeAl$_2$, the observed ``double-q'' state\cite{CEAL5,CEAL6} is
favored over any other combination of wavevector(s) in the
star of ${\bf q}$.  This result is not a common one for a cubic system.
In addition our analysis reproduces the relative phase observed\cite{CEAL5}
between the two coupled wavevectors.

\item
By relating the exchange constants to the Curie-Weiss intercept temperature
$\Theta_{\rm C-W}$ of the inverse susceptibility and to the ordering
temperature, we developed the estimates for the nearest neighbor
ferromagnetic interaction, $K/k=-11\pm 1$K, and for the next-nearest neighbor
antiferromagnetic interaction, $J/k=6 \pm 1$K.

\item
We also showed (see appendix C) that the exchange interactions are
significantly renormalized by virtual crystal field excitations.
This effect leads to rather long-range exchange interactions.

\item
It is possible that our analysis of wavevector selection can explain the
similar incommensurate wavevector observed\cite{TMS2,TMS3} for the Kondo-like
system TmS.  Although the anisotropy axis  is different for TmS than for CEAL,
one may speculate that the fourth order terms in TmS may give rise
to a double-q state, although such a state has not yet been observed in TmS.
\end{itemize}


\vspace{0.2 in}
\begin{appendix}
\section{SPIN FUNCTIONS FOR THE {\bf STAR OF q}}

In this appendix we determine the spin functions for the different
wavevectors in the star of ${\bf q}$, given that for
\begin{eqnarray}
a{\bf q}/(2 \pi) = 1/2-\delta, 1/2+\delta, 1/2
\label{QVEC} \end{eqnarray}
the spin functions for the two sites in the unit cell are\cite{CEAL2,CEAL5}
\begin{eqnarray}
{\bf m}_1({\bf q}) &=& [ \alpha e^{i\phi} , \alpha e^{-i\phi} \ , \beta ]
= {\bf m}_1 (-{\bf q})^*
\nonumber \\
{\bf m}_2({\bf q}) &=& [ - \alpha e^{-i\phi} , - \alpha e^{i\phi} \ , -\beta ]
\ , = {\bf m}_2 (-{\bf q})^*
\label{ALPHAEQ} \end{eqnarray}
where $\alpha$, $\beta$, and $\phi$ are real valued constants.
are fixed by the interactions through the quadratic terms in the
free energy.  Since we will study the quartic terms which couple
different wavevectors, we need to tabulate the spin functions for
the different wavevectors.


The star of the wavevector consists of 24 vectors which are
$\pm {\bf Q}^\alpha_n$, $\pm {\bf Q}^\beta_n$, and
$\pm {\bf Q}^\gamma_n$, for $n=1, 2, 3, 4$. These ${\bf Q}$'s
are listed in Tables \ref{WAVE1}, \ref{WAVE2}, and \ref{WAVE3}.
The spin functions for different
wavevectors are related by the symmetry operations of the crystal,
which is space group \#227, Fd$\overline 3$m, in the International
Tables for Crystallography (ITC).\cite{HAHN}

In Eq. (\ref{ALPHAEQ}) we gave the spin wavefunction for
${\bf Q}^\alpha_1$.  We now consider the effect on this function
of the operation $(x,y,z) \rightarrow (-y, -x, z)$
(\#37 in ITC), which we regard as
a mirror which interchanges $x$ and $y$ followed by a two-fold
rotation about $z$.  Because spin is a pseudovector this operation
on spin is
\begin{eqnarray}
(m_x, m_y, m_z) \rightarrow (m_y, m_x, -m_z) \ .
\label{MPRIME} \end{eqnarray}
Thus, before transformation we have
\begin{eqnarray}
m_x({\bf R}_i, \tau_1) &=&
2\alpha \cos({\bf q}_i \cdot {\bf R}_i + \phi) \nonumber \\
m_y({\bf R}_i, \tau_1) &=&
2\alpha \cos({\bf q}_i \cdot {\bf R}_i - \phi) \nonumber \\
m_z({\bf R}_i, \tau_1) &=&
2\beta \cos({\bf q}_i \cdot {\bf R}_i ) \nonumber \\
m_x({\bf R}_i, \tau_2) &=&
-2\alpha \cos({\bf q}_i \cdot {\bf R}_i - \phi) \nonumber \\
m_y({\bf R}_i, \tau_2) &=&
-2\alpha \cos({\bf q}_i \cdot {\bf R}_i + \phi) \nonumber \\
m_z({\bf R}_i, \tau_2) &=&
-2\beta \cos({\bf q}_i \cdot {\bf R}_i ) \ ,
\label{BEFORE} \end{eqnarray}
where ${\bf R}_i \equiv (X_i,Y_i,Z_i)$ specifies the location of the unit cell
before transformation, ${\bf q}_i$ is the wavevector
before transformation, given in Eq. (\ref{QVEC}), and
\begin{eqnarray}
\tauv_1 = (0,0,0) \ , \ \ \ \ 
\tauv_2 = a(1,1,1)/4 \ .
\end{eqnarray}
After transformation (indicated by primes) Eq. (\ref{MPRIME}) gives
\begin{eqnarray}
m_x^\prime({\bf R}_f, \tau_{1,f}) &=&
2\alpha \cos({\bf q}_i \cdot {\bf R}_i - \phi) \nonumber \\
m_y^\prime({\bf R}_f, \tau_{1,f}) &=&
2\alpha \cos({\bf q}_i \cdot {\bf R}_i + \phi) \nonumber \\
m_z^\prime({\bf R}_f, \tau_{1,f}) &=&
-2\beta \cos({\bf q}_i \cdot {\bf R}_i ) \nonumber \\
m_x^\prime({\bf R}_f, \tau_{2,f}) &=&
-2\alpha \cos({\bf q}_i \cdot {\bf R}_i + \phi) \nonumber \\
m_y^\prime({\bf R}_f, \tau_{2,f}) &=&
-2\alpha \cos({\bf q}_i \cdot {\bf R}_i - \phi) \nonumber \\
m_z^\prime({\bf R}_f, \tau_{2,f}) &=&
2\beta \cos({\bf q}_i \cdot {\bf R}_i ) \ .
\end{eqnarray}
where ${\bf R}_f= (X_f,Y_f,Z_f)$. If the initial position is
\begin{eqnarray}
{\bf r}&=& (X_i,Y_i,Z_i) + \tauv_1 = (X_i,Y_i,Z_i) \ ,
\end{eqnarray}
then the final position is
\begin{eqnarray}
{\bf r}^\prime &=& (-Y_i,-X_i,Z_i) \equiv (X_f,Y_f,Z_f) + \tauv_f \ ,
\end{eqnarray}
so that $\tauv_{1f} = \tauv_1$.  We now
express ${\bf q}_i \cdot {\bf R}_i$ in terms of the final coordinates:
\begin{eqnarray}
q_{ix}X_i + q_{iy}Y_i + q_{iz}Z_i &=& q_{ix} Y_f - q_{iy}X_f + q_{iz}Z_f \ ,
\end{eqnarray}
which can be written as ${\bf q}_i \cdot {\bf R}_i
={\bf q}_f \cdot {\bf R}_f$, where we have
(to within a reciprocal lattice vector)
\begin{eqnarray}
{\bf q}_f &=& (-q_{i,y}, -q_{i,x},q_{i,z}) =
{\bf Q}^\alpha_2 \ .
\end{eqnarray}
Thus
\begin{eqnarray}
m_x^\prime({\bf R}_f, \tau_1) &=&
2\alpha \cos({\bf Q}^\alpha_2 \cdot {\bf R}_f - \phi) \nonumber \\
m_y^\prime({\bf R}_f, \tau_1) &=&
2\alpha \cos({\bf Q}^\alpha_2 \cdot {\bf R}_f + \phi) \nonumber \\
m_z^\prime({\bf R}_f, \tau_1) &=&
-2\beta \cos({\bf Q}^\alpha_2 \cdot {\bf R}_f ) \ .
\end{eqnarray}
Now consider $\tauv_i=\tauv_2$. Then if the initial position is
\begin{eqnarray}
{\bf r}&=& (X_i+a/4,Y_i+a/4,Z_i+a/4) \ ,
\end{eqnarray}
then the final position is
\begin{eqnarray}
{\bf r}^\prime &=& (-Y_i-a/4,-X_i-a/4,Z_i+a/4) 
\nonumber \\ &\equiv& (X_f,Y_f,Z_f) + \tauv_f \ ,
\end{eqnarray}
so that, in this case, $\tauv_f=\tauv_2$.
\begin{eqnarray}
X_f= -Y_i-a/2 \ , \ \ Y_f= -X_i-a/2 \ , \ \ Z_f= Z_i \ .
\end{eqnarray}
We express ${\bf q}_i \cdot {\bf R}_i$ in terms of the final coordinates:
\begin{eqnarray}
{\bf q}_i \cdot {\bf R}_i &=&
q_{ix}(-Y_f-a/2) + q_{iy}(-X_f-a/2) + q_{iz}Z_i \nonumber \\
&=& {\bf Q}_2^\alpha \cdot {\bf R}_f -  \pi \ .
\end{eqnarray}
Thus
\begin{eqnarray}
m_x^\prime({\bf R}_f, \tau_2) &=&
2\alpha \cos({\bf Q}^\alpha_2 \cdot {\bf R}_f + \phi) \nonumber \\
m_y^\prime({\bf R}_f, \tau_2) &=&
2\alpha \cos({\bf Q}^\alpha_2 \cdot {\bf R}_f - \phi) \nonumber \\
m_z^\prime({\bf R}_f, \tau_2) &=&
-2\beta \cos({\bf Q}^\alpha_2 \cdot {\bf R}_f ) \ .
\label{TRANSEQ} \end{eqnarray}
Thus for wavevector ${\bf Q}^\alpha_2$ the Fourier component vector (which we 
put into Table \ref{WAVE1}) is
\begin{eqnarray}
( \alpha e ^{-i \phi} \ , \ \alpha e ^{i \phi} \ , \ -\beta \ ;
\ \alpha e ^{i \phi} \ , \ \alpha e ^{-i \phi} \ , \ - \beta ) \ .
\end{eqnarray}
Next we study the effect of the transformation $(x,y,z) \rightarrow
(x+1/4,-y+1/4,z+1/4)$.  Before transformation the Fourier
coefficients are those of Eq. (\ref{BEFORE}).  Since this transformation
is a mirror operation we have, after transformation that
\begin{eqnarray}
m_x^\prime({\bf R}_f, \tau_{1f}) &=&
-2\alpha \cos({\bf q}_i \cdot {\bf R}_i + \phi) \nonumber \\
m_y^\prime({\bf R}_f, \tau_{1f}) &=&
2\alpha \cos({\bf q}_i \cdot {\bf R}_i - \phi) \nonumber \\
m_z^\prime({\bf R}_f, \tau_{1f}) &=&
-2\beta \cos({\bf q}_i \cdot {\bf R}_i ) \nonumber \\
m_x^\prime({\bf R}_f, \tau_{2f}) &=&
2\alpha \cos({\bf q}_i \cdot {\bf R}_i - \phi) \nonumber \\
m_y^\prime({\bf R}_f, \tau_{2f}) &=&
-2\alpha \cos({\bf q}_i \cdot {\bf R}_i + \phi) \nonumber \\
m_z^\prime({\bf R}_f, \tau_{2f}) &=&
2\beta \cos({\bf q}_i \cdot {\bf R}_i ) \ .
\end{eqnarray}
For $\tau_i=1$ the initial position is
\begin{eqnarray}
{\bf r} &=& {\bf R}_i + \tau_1 = (X_i,Y_i,Z_i) ,
\end{eqnarray}
and, using the transformation, the final position is
\begin{eqnarray}
{\bf r}^\prime &=& (X_i + a/4, -Y_i+a/4, Z_i+a/4) \nonumber \\
&\equiv& (X_f, Y_f, Z_f) + \tau_{1,f} \ .
\end{eqnarray}
Thus $\tau_{1f}=\tauv_2$ and
\begin{eqnarray}
{\bf q}_i \cdot {\bf R}_i &=& {\bf Q}_3^\alpha \cdot {\bf R}_f \ .
\end{eqnarray}
Then
\begin{eqnarray}
m_x^\prime({\bf R}_f, \tau_{2}) &=&
-2\alpha \cos({\bf Q}^\alpha_3 \cdot {\bf R}_f + \phi) \nonumber \\
m_y^\prime({\bf R}_f, \tau_{2}) &=&
2\alpha \cos({\bf Q}^\alpha_3 \cdot {\bf R}_f - \phi) \nonumber \\
m_z^\prime({\bf R}_f, \tau_{2}) &=&
-2\beta \cos({\bf Q}^\alpha_3 \cdot {\bf R}_f ) \ .
\label{T3A} \end{eqnarray}
Using the transformation on ${\bf r} \equiv (X_i+a/4,Y_i+a/4,Z_i+a/4)$, 
we write
\begin{eqnarray}
{\bf r}' &=& (X_i+a/2, -Y_i, Z_i+a/2) \nonumber \\ 
&\equiv& (X_i+a/2, -Y_i, Z_i+a/2) + \tauv_1 \ ,
\end{eqnarray}
so that $\tau_{2f}=\tau_1$ and
\begin{eqnarray}
{\bf R}_f &=& (X_i+a/2,-Y_i,Z_i+a/2) \ .
\end{eqnarray}
Thus
\begin{eqnarray}
{\bf q}_i \cdot {\bf R}_i &=& {\bf q}_f \cdot {\bf R}_f 
- {\bf q}_f \cdot (a/2,0,a/2) \nonumber \\ &=&
{\bf Q}^\alpha_3 \cdot {\bf R}_f - \pi + \pi \delta \ ,
\end{eqnarray}
so that
\begin{eqnarray}
m_x^\prime({\bf R}_f, \tau_{1}) 
&=& -2\alpha \cos({\bf Q}^\alpha_3 \cdot {\bf R}_f - \phi + \pi \delta )
\nonumber \\
m_y^\prime({\bf R}_f, \tau_{1}) 
&=& 2\alpha \cos({\bf Q}^\alpha_3 \cdot {\bf R}_f + \phi + \pi \delta )
\nonumber \\
m_z^\prime({\bf R}_f, \tau_{1}) 
&=& -2\beta \cos({\bf Q}^\alpha_3 \cdot {\bf R}_f  + \pi \delta ) \ .
\label{T3B} \end{eqnarray}
Thus for wavevector ${\bf Q}^\alpha_3$ the Fourier component vector (which we 
put into Table \ref{WAVE1}) is
\begin{eqnarray}
(&&-\alpha e ^{i (\pi \delta - \phi)} \ , \ \alpha e ^{i(\pi \delta + \phi )}
\ , \ -\beta e^{i \pi \delta} \ ; \nonumber \\ &&  - \alpha e ^{i\phi } \ ,
\ \alpha e ^{-i \phi } \ , \ - \beta ) \ .
\end{eqnarray}

Next we study the effect of the transformation $(x,y,z) \rightarrow
(-y+1/4,x+1/4,z+1/4)$ (\#16 in ITC). Since this transformation is a
four-fold screw axis, we have, after transformation that
\begin{eqnarray}
m_x^\prime({\bf R}_f, \tau_{1f}) &=&
-2\alpha \cos({\bf q}_i \cdot {\bf R}_i - \phi) \nonumber \\
m_y^\prime({\bf R}_f, \tau_{1f}) &=&
2\alpha \cos({\bf q}_i \cdot {\bf R}_i + \phi) \nonumber \\
m_z^\prime({\bf R}_f, \tau_{1f}) &=&
2\beta \cos({\bf q}_i \cdot {\bf R}_i ) \nonumber \\
m_x^\prime({\bf R}_f, \tau_{2f}) &=&
2\alpha \cos({\bf q}_i \cdot {\bf R}_i + \phi) \nonumber \\
m_y^\prime({\bf R}_f, \tau_{2f}) &=&
-2\alpha \cos({\bf q}_i \cdot {\bf R}_i - \phi) \nonumber \\
m_z^\prime({\bf R}_f, \tau_{2f}) &=&
- 2 \beta \cos({\bf q}_i \cdot {\bf R}_i ) \ .
\end{eqnarray}
For $\tau_i=\tau_1$, and if ${\bf R}_i =(X_i,Y_i,Z_i)$, we have
\begin{eqnarray}
{\bf r}^\prime &=& 
(-Y_i+a/4,X_i+a/4,Z_i+a/4) \ ,
\end{eqnarray}
so that $\tauv_{1f}=\tauv_2$ and to within a reciprocal lattice vector
this gives
\begin{eqnarray}
{\bf q}_f &=& {\bf Q}^\alpha_4 \ ,
\end{eqnarray}
so that
\begin{eqnarray}
m_x^\prime({\bf R}_f, \tau_{2}) &=&
-2\alpha \cos({\bf Q}^\alpha_4 \cdot {\bf R}_f - \phi) \nonumber \\
m_y^\prime({\bf R}_f, \tau_{2}) &=&
2\alpha \cos({\bf Q}^\alpha_4 \cdot {\bf R}_f + \phi) \nonumber \\
m_z^\prime({\bf R}_f, \tau_{2}) &=&
2\beta \cos({\bf Q}^\alpha_4 \cdot {\bf R}_f ) \ .
\end{eqnarray}
For $\tau_i=\tau_2$, ${\bf r} = (X_i+a/4, Y_i+a/4, Z_i+a/4)$ and
\begin{eqnarray}
{\bf r}^\prime &=& (-Y_i,X_i+a/2,Z_i+a/2) \ ,
\end{eqnarray}
so that $\tauv_{2f}=\tauv_1$ and
\begin{eqnarray}
{\bf q}_i \cdot {\bf R}_i 
&=& {\bf Q}^\alpha_4 \cdot {\bf R}_f - (a/2)(q_{xi}+q_{zi}) \nonumber \\ 
&=& {\bf Q}^\alpha_4 \cdot {\bf R}_f + \pi (-1 + \delta) \ .
\end{eqnarray}
so that
\begin{eqnarray}
m_x^\prime({\bf R}_f, \tau_{1}) &=& -2\alpha
\cos({\bf Q}^\alpha_4 \cdot {\bf R}_f + \phi + \pi \delta ) \nonumber \\
m_y^\prime({\bf R}_f, \tau_{1}) &=&
2\alpha \cos({\bf Q}^\alpha_4 \cdot {\bf R}_f - \phi + \pi \delta )
\nonumber \\
m_z^\prime({\bf R}_f, \tau_{1}) &=&
2\beta \cos({\bf Q}^\alpha_4 \cdot {\bf R}_f  + \pi \delta ) \ .
\end{eqnarray}

Thus we have the results for the order parameter wavefunctions given in
Table \ref{WAVE1}.  To get the wavefunctions for ${\bf Q}^\beta_n$ and
for ${\bf Q}^\gamma_n$
is much easier: one simply uses the three-fold rotation axis about (111)
to get the results given in Tables \ref{WAVE2} and \ref{WAVE3}.

\section{CURIE-WEISS SUSCEPTIBILITY IN A CRYSTAL FIELD}

\begin{figure}
\begin{center}
\includegraphics[width=8cm]{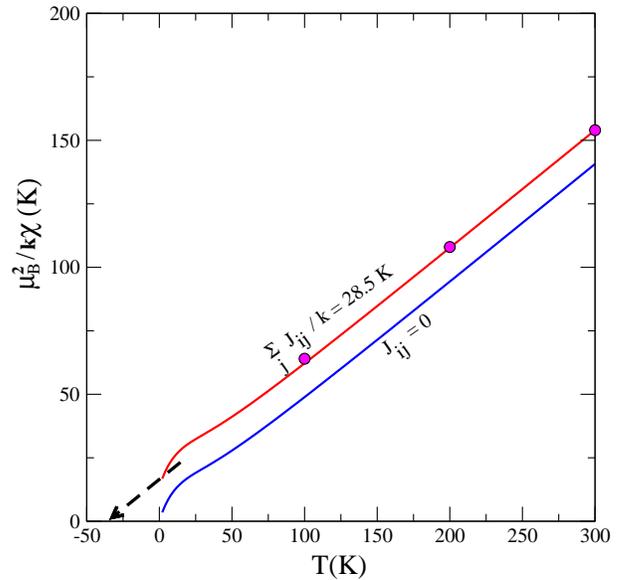} 
\caption{\label{CHIEPS}
(Color online) Inverse susceptibility, $1/\chi$.  The dots are data points taken
from the more extensive data set of Ref. 35.  The lower (online blue)
curve is for $\lambda=0$ and the upper (online red) one is correct to first
order in $\lambda$ [using Eq. (\protect{\ref{LOCAL}})]
for $\sum_j J_{ij}/k=28.5K$. For this curve the
intercept extrapolated from $280{\rm K}<T<300{\rm }$ is $-33$K, as indicated
by the arrow. The intercept extrapolated from infinite temperature
is $\theta_{\rm C-W} = -(21/20) \sum_j J_{ij}=-30$K. So even at $T=300$K
there is still a noticeable departure [of order $(T_Q/T)^2$] from the infinite
temperature behavior.}
\end{center}
\end{figure}

Here we develop a formula for the susceptibility correct to
leading order in the exchange interactions, $J_{ij}$.  For this
purpose we write the Hamiltonian as
\begin{eqnarray}
{\cal H} &=& {\cal H}_0 + \lambda \sum_{i<j} J_{ij}
{\bf S}_{{\rm eff},i} \cdot {{\bf S}_{\rm eff},j} \ ,
\end{eqnarray}
where the ${\bf S}_{{\rm eff},i}$ is the effective spin 1/2 operator
we have used throughout our calculations and $\lambda$, a scale
factor for the perturbation, is set equal to unity in the final
results.  Here ${\cal H}_0$ includes
all terms for $J_{ij}=0$.  Thus ${\cal H}_0$ is the
Hamiltonian for spins subject to the cubic crystal field
and the external magnetic field, but with no exchange interactions between
neighboring spins.  It will be convenient to express this Hamiltonian
in terms of the magnetic moment operator $\muv_i$ for site $i$.
We write ${\bf S}_{\rm eff}= (3/5){\bf J} = [3/(5g_J\mu_B)]\muv$, so that
(with $g_J=6/7$) 
\begin{eqnarray}
{\cal H} &=& {\cal H}_0 + (7/10\mu_B)^2 \lambda \sum_{i<j} J_{ij}
\muv_i \cdot \muv_j \ ,
\end{eqnarray}
Correct to leading order in $\lambda$ we use thermodynamic perturbation
theory\cite{LL} to write the free energy as
\begin{eqnarray}
F(\lambda) &=& F(\lambda =0) + {49 \lambda \over 100\mu_B^2} 
\sum_{i<j} J_{ij} \langle \muv_i \rangle_0 \langle \muv_j \rangle_0 \ ,
\end{eqnarray}
where $F(\lambda =0)$ is the free energy for the Hamiltonian ${\cal H}_0$ and
\begin{eqnarray}
\langle X \rangle_0 & \equiv & {\rm Tr} [X e^{-\beta {\cal H}_0}] /
{\rm Tr} [e^{-\beta {\cal H}_0}] \ .
\end{eqnarray}
Then the susceptibility per spin, $\chi \equiv \partial \langle \mu_i
\left. \rangle / \partial H \right|_{H=0}$, is
\begin{eqnarray}
\chi(\lambda) &=& N^{-1} {\partial^2 F(\lambda) \over \partial H^2 } =
\chi (\lambda=0) \nonumber \\ && \
- [49 \lambda /(100 \mu_B^2)] \sum_j J_{ij} \chi(\lambda=0)^2 \ ,
\end{eqnarray}
where $N$ is the total number of Ce ions.  Thus
\begin{eqnarray}
\chi(\lambda)^{-1}&=& \chi(\lambda=0)^{-1} + [49 \lambda /(100 \mu_B^2)]
\sum_j J_{ij} + {\cal O} (\lambda^2) \nonumber \\ &=& \chi(\lambda=0)^{-1}
+ [49 /(100 \mu_B^2)](4K+12J + \dots ) \nonumber \\ && \
+ {\cal O} (J_{ij}^2) \ .
\label{LOCAL} \end{eqnarray}
To obtain $\chi(\lambda=0)$ we took the wavefunctions of the ground doublet
in the cubic crystal field to be
\begin{eqnarray}
|0\rangle_\pm &=& \sqrt {5/6} |5/2, \pm 3/2\rangle
- \sqrt{1/6} |5/2, \mp 5/2 \rangle\ ,
\label{WAVEFN} \end{eqnarray}
in the $J,J_z$ representation.  The remaining states form the four-fold
degenerate excited state at a relative energy which we denote $kT_Q$.
Then we found that
\begin{eqnarray}
{k \chi(\lambda=0) \over \mu_B^2}
&=& {320 (1 - e^{-T_Q/T}) \over 49T_Q[ 2+4e^{-T_Q/T}]} \nonumber \\ && \
+ {50+260e^{-T_Q/T} \over 49T[2 + 4e^{-T_Q/T}]} \ .
\end{eqnarray}
At low temperature, the second term displays the Curie-like $1/T$
dependence corresponding to the moment in the ground doublet and
the first term is the so-called Van Vleck temperature-independent
susceptibility.\cite{VV}
To illustrate the effect of this term, we show the inverse susceptibility
in Fig. \ref{CHIEPS} for the Ce ion ($J=5/2)$ in a cubic crystal field
with a doublet-quartet energy splitting of $kT_Q$ with $T_Q=100$K.
In the high temperature limit $T>>T_Q$ we have
\begin{eqnarray}
\chi(\lambda=0) &=& {g_J^2 J(J+1) \mu_B^2 \over 3kT} = {15 \mu_B^2 
\over 7 kT} \ ,
\end{eqnarray}
in which case for $\lambda=1$ we have
\begin{eqnarray}
\chi^{-1} = {7 k \over 15 \mu_B^2 } \left[ T - \Theta_{\rm C-W} \right] 
+ {\cal O}\left\{ J_{ij}/T, (T_Q/T)^2 \right\} \ ,
\end{eqnarray}
with
\begin{eqnarray}
\Theta_{\rm C-W} &=& - 21 \sum_j J_{ij} / 20 \approx - 21(K+3J)/5 \ .
\end{eqnarray}

In Fig. \ref{CHIEPS} we also show the inverse susceptibility when
$\Theta_{{\rm C} - {\rm W}}=-29.9$K, a value which gives the
Curie-Weiss intercept (extrapolated from $T=300$) of
$-33$K as in Ref. \onlinecite{CURIE1}.

\section{EFFECTIVE INTERACTIONS VIA EXCITED QUARTET STATES}

Here we consider effective interactions which occur via excited
virtual crystal field states.  In a general formulation one
considers manifolds ${\cal M}_n$ in which $n$ spins are in their
excited quartet crystal field level, whose energy is $kT_Q$ relative
to the crystal field ground state.  We are interested in the
effective Hamiltonian ${\cal H}_0$ for ${\cal M}_0$ at low
temperature and here we discuss its evaluation within low order
perturbation theory.  We define
\begin{eqnarray}
{\cal H}_{n,m} &=& {\cal P}_n {\cal H} {\cal P}_m \ ,
\end{eqnarray}
where ${\cal P}_n$ is the projection operators for the manifold
${\cal M}_n$.  Clearly the lowest approximation is to neglect entirely
all processes except those within the manifold ${\cal M}_0$ and this
was an implicit assumption of our calculations in the body of this paper.
Processes involving virtual state in the manifold ${\cal M}_1$ enter via
second order perturbation theory.  These terms are obtained
just as for superconductivity,\cite{CK} with the result that
\begin{eqnarray}
{\cal H}_0 &=& {\cal H}_{0,0} - {1 \over kT_Q} \sum_n n^{-1}
{\cal H}_{0,n} {\cal H}_{n,0} \ .
\label{OFF} \end{eqnarray}
One can use this formalism to reproduce the formula for the zero-temperature
Van Vleck susceptibility tensor,\cite{VV}
$\chi^{(V)}_{\alpha ,\beta}$.  However,
our present aim is rather to analyze effective exchange interactions
which arise in this way.  To illustrate the phenomenon, consider
contributions to Eq. (\ref{OFF}) when ${\cal H}_{0,n}$ is taken
to be the nn exchange interaction.  Although we wrote this interaction
as $K_0{\bf S}_{\rm eff,i} \cdot {\bf S}_{\rm eff,j}$ it really should
be represented as $K'{\bf J}_i \cdot {\bf J}_j$, where, since
${\bf S}_{\rm eff ,i}= [(g_J-1)/g_0]{\bf J}_i=3/5 {\bf J}_i$ one
has that $K'=(3/5)^2K_0=(9/25)K_0$. Then we obtain a contribution $V_{ij}$
to the effective nnn exchange interaction between spins $i$ and $j$ 
[at separation (a/2,a/2,0)] using
the nn interactions $J_{ik}$ between spins $i$ and $k$ and $J_{kj}$
between $k$ and $j$. Since for a nnn pair $i,j$ there is only one choice
for the intermediate site $k$ to be an nn of both sites $i$ and $j$,
we have
\begin{eqnarray}
V_{ij} &=& = -{{K'}^2 \over kT_Q} {\cal P}_0 {\bf J}_i \cdot {\bf J}_k
{\cal P}_1 {\bf J}_k \cdot {\bf J}_j {\cal P}_0 \ .
\label{DELJ} \end{eqnarray}
Because of the cubic symmetry of the crystal field one has
\begin{eqnarray}
{\cal P}_0 J_{k,\alpha} {\cal P}_1 J_{k,\beta} {\cal P}_0
=  (20/9) \delta_{\alpha , \beta} {\cal P}_0 \ .
\label{PP2EQ} \end{eqnarray}
To obtain this result it is convenient to take $\alpha=\beta=z$
and use the ground state wavefunctions of Eq. (\ref{WAVEFN}).
Then Eq. (\ref{DELJ}) yields
\begin{eqnarray}
V_{ij} &=&  -{20(9K_0/25)^2 \over 9kT_Q} {\cal P}_0 {\bf J}_i \cdot
{\bf J}_j {\cal P}_0 \nonumber \\ &=&
- {4K_0^2 \over 5 kT_Q} {\bf S}_{\rm eff,i} \cdot {\bf S}_{\rm eff,j} \ .
\end{eqnarray}
This means that due to these processes we have that
\begin{eqnarray}
J \rightarrow  J - {4K_0^2 \over 5 kT_Q} \equiv J - \delta J \ .
\end{eqnarray}
For $K_0/k=10$K and $T_Q=100$K, this gives $\delta J/k \approx 0.8$K.
The nn interaction is renormalized in a similar way, except that
if sites $i$ and $j$ are nn's, then there are now two choices
for the site $k$ to be an nn of both $i$ and $j$.  Thus
\begin{eqnarray}
K_0 \rightarrow K_0 - {8K_0^2 \over 5kT_Q} \equiv K_0 - \delta K_0 \ ,
\end{eqnarray}
with $\delta K_0 = 1.6$K.  As a final example, we similarly find 
for sites separated by $(a,0,0)$ that there are
four intermediate paths of sites separated by $(a/2,a/2,0)$, so that
\begin{eqnarray}
M \rightarrow M- {16 J^2 \over 5kT_Q} \equiv M - \delta M \ .
\end{eqnarray}
Taking $J/k=5$K, we find that
$\delta M/k \approx 0.8$K, so that $\delta M/J = 0.16$, 
a value which is comparable to those used in Table  \ref{EXTABLE}.

These results imply that even if the bare
Hamiltonian only has nn interactions, virtual processes involving
higher crystal field states will induce nnn interactions approximately
of the size we will deduce from fitting experiments.  This mechanism
in higher order will produce significant longer range interactions
even if the bare Hamiltonian has only nn interactions initially.
\end{appendix}


\begin{thebibliography} {99}
\bibitem{FN}
For general spin $S$ the factor $4kT$ in Eq. (\ref{CHIVEQ}) would be
$3kT/[S(S+1)]$ and would reproduce the correct $T_c$ for, say,
a simple cubic ferromagnet.
\bibitem{LOCK}
The minimum for ${\cal S}$ occurs when only $x_{2n-1}$ and $x_{2n}$
are nonzero in which case ${\cal S}$ assumes the value given by the
right-hand side of Eq.
(\ref{S12EQ}).  Then, if $(x_{2n-1},x_{2n})$ minimizes ${\cal S}$,
so also does $(e^{i \eta} x_{2n-1},e^{i \eta} x_{2n})$. Thus, at this
level, the overall phase $\eta$ is not locked.  The overall phase
is locked by terms of $p$th order in the $x$'s, such that the condition
$p {\bf q}_i={\bf G}$ is very nearly satisfied.
\end{thebibliography}
\end{document}